\documentclass[aps,prd,groupedaddress,floatfix,nofootinbib,notitlepage]{revtex4-1}

\usepackage[a4paper, total={6.83in, 9.47in}]{geometry}

\usepackage[utf8x]{inputenc}
\usepackage{graphicx}
\usepackage{hyperref}
\usepackage{latexsym}
\usepackage{amsmath}
\usepackage{amssymb}
\usepackage{pdfsync}
\usepackage{slashed}
\usepackage{mathrsfs}
\usepackage[math]{cellspace}
\usepackage{multirow}
\usepackage{hhline}
\usepackage{xcolor} 
\newcommand{\be}{\begin{equation}}
\newcommand{\ee}{\end{equation}}
\newcommand{\ba}{\begin{array}}
\newcommand{\ea}{\end{array}}
\newcommand{\bea}{\begin{equation}\begin{aligned}}
\newcommand{\eea}{\end{aligned}\end{equation}}
\def\lsim{\mathrel{\raise.3ex\hbox{$<$\kern-.75em\lower1ex\hbox{$\sim$}}}}
\def\gsim{\mathrel{\raise.3ex\hbox{$>$\kern-.75em\lower1ex\hbox{$\sim$}}}}
\newcommand{\bear}{\begin{eqnarray}}
\newcommand{\eear}{\end{eqnarray}}
\newcommand{\sss}{\scriptscriptstyle}
\newcommand{\W}{{\sss W}}

\long\def\/*#1*/{}

\begin{document}
	\title{Baryogenesis and gravity waves from a 
UV-completed\\ electroweak phase transition}

\author{Benoit Laurent}
\email{benoit.laurent@mail.mcgill.ca}
\affiliation{McGill University, Department of Physics, 3600 University St.,
Montr\'eal, QC H3A2T8 Canada}
\author{Avi Friedlander}
\email{avi.friedlander@queensu.ca}
\affiliation{Queen's University, Department of Physics \& Engineering Physics Astronomy
Kingston, Ontario, K7L 3N6 Kingston, Canada}
\author{Dong-Ming He}
\email{dong-ming.he@student.uva.nl}
\affiliation{University of Science and Technology of China, Hefei, Anhui
230026}
\affiliation{Universiteit van Amsterdam, Science Park 904, Amsterdam, 1098XH,
Netherlands}
\author{James M.~Cline}
\email{jcline@physics.mcgill.ca}
\affiliation{McGill University, Department of Physics, 3600 University St.,
Montr\'eal, QC H3A2T8 Canada}
\author{Kimmo Kainulainen}
\email{kimmo.kainulainen@jyu.fi}
\affiliation{Department of Physics, P.O.Box 35 (YFL), 
             FIN-40014 University of Jyv\"askyl\"a, Finland}
\affiliation{Helsinki Institute of Physics, P.O. Box 64,
             FIN-00014 University of Helsinki, Finland}
\author{David Tucker-Smith}
\email{dtuckers@williams.edu}
\affiliation{Department of Physics, Williams College, Williamstown, MA 01267}

\begin{abstract}

We study gravity wave production and baryogenesis at the electroweak phase transition, in a real singlet scalar extension of the Standard Model, including vector-like top partners to generate the CP violation needed for electroweak baryogenesis (EWBG). The singlet makes the phase transition strongly first-order through its coupling to the Higgs boson, and it spontaneously breaks CP invariance through a dimension-5 contribution to the top quark mass term, generated by integrating out the heavy top quark partners. We improve on previous studies by incorporating updated transport equations, compatible with large bubble wall velocities. 
The wall speed and thickness are computed directly from the microphysical parameters rather than treating them as free parameters, allowing for a first-principles computation of the baryon asymmetry. The size of the CP-violating dimension-5 operator needed for EWBG is constrained by collider, electroweak precision, and renormalization group running constraints. We identify regions of parameter space that can produce the observed baryon asymmetry or observable gravitational (GW) wave signals. Contrary to standard lore, we find that for strong deflagrations, the efficiencies of large baryon asymmetry production and strong GW signals can be {\em positively} correlated. 
However we find the overall likelihood of observably large GW signals to be smaller than estimated in previous studies.  In particular, only detonation-type transitions are predicted to produce observably large gravitational waves.

\end{abstract}
\maketitle
\tableofcontents

%
\section{Introduction}
\label{sec:intro}
%

Phase transitions in the early universe provide an opportunity for probing physics at high scales through cosmological observables, in particular, if the transition is first order.
In that case, it may be possible to explain the origin of baryonic matter through electroweak baryogenesis (EWBG) \cite{Bochkarev:1990fx,Cohen:1990py,Cohen:1990it,Turok:1990zg} or variants thereof \cite{Long:2017rdo}.  Such transitions can also produce relic gravitational waves (GWs) that may be detectable by future experiments like LISA \cite{caprini2016science,Caprini:2019egz}, 
BBO \cite{Crowder:2005nr}, DECIGO \cite{Seto:2001qf,Sato:2017dkf} and AEDGE \cite{Bertoldi:2019tck}.  \\

It is remarkable that even though the electroweak phase transition (EWPT) is a smooth crossover in the standard model (SM)
\cite{Kajantie:1996qd,Kajantie:1996mn}, it can become first order with the addition of modest new physics input, in particular a singlet scalar coupling to the Higgs  \cite{Anderson:1991zb,McDonald:1993ey,Choi:1993cv,Espinosa:2007qk,Profumo:2007wc,Espinosa:2011ax,Cline:2013gha,Damgaard:2015con,Kurup:2017dzf,Chiang:2018gsn}, that can also be probed in collider experiments \cite{Ham:2004cf,Noble:2007kk,Curtin:2014jma,Profumo:2014opa,Kotwal:2016tex,Huang:2016cjm,Chen:2017qcz,Kim:2018mks,Hashino:2018wee,Ramsey-Musolf:2019lsf,Xie:2020wzn,Adhikari:2020vqo}.  There have been many studies of such new physics models with respect to their potential to produce observable cosmological signals \cite{Das:2009ue,Ashoorioon:2009nf,Kakizaki:2015wua,Hashino:2016rvx,Chala:2016ykx,Tenkanen:2016idg,Hashino:2016xoj,vaskonen2017electroweak,Beniwal:2017eik,Ahriche:2018rao,DeCurtis:2019rxl,beniwal2019gravitational,Carena:2019une,Ellis:2020nnr}.  However, it is challenging to make a first-principles connection between microphysical models and the baryon asymmetry or GW production, since these can be sensitive to the velocity $v_w$ and thickness $L_w$ of the bubble walls in the phase transition, which are numerically demanding to compute \cite{Moore:1995si,Moore:1995ua,John:2000zq,Bodeker:2009qy,Kozaczuk:2015owa,
Konstandin:2010dm,Konstandin:2014zta,Bodeker:2017cim,Hoeche:2020rsg,Mancha:2020fzw,Balaji:2020yrx,Vanvlasselaer:2020niz}.
Most previous studies that encompass EWBG and GW studies of the EWPT therefore leave $v_w$ and $L_w$ as free parameters.  This limitation was addressed recently in Ref.\ \cite{Friedlander:2020tnq}, which undertook a comprehensive investigation of the EWPT enhanced by coupling the Higgs boson to a scalar singlet with $Z_2$ symmetry.  The simplicity of this model facilitates doing an exhaustive search of its parameter space.\\

In the present work we continue the investigation started in Ref.\  \cite{Friedlander:2020tnq}, which determined $v_w$ and $L_w$ over much of the model parameter space, but did not try to predict the baryon asymmetry or GW production.  Moreover, that study was limited to subsonic wall speeds, due to a breakdown of the fluid equations that determine the friction on the wall.  Recently a set of improved fluid equations was postulated in Refs.\ \cite{Cline:2020jre,Laurent:2020gpg}, that do not suffer from the subsonic limitation.  We use these in the present work in order to fully explore the parameter space, where high $v_w$ can be favorable to observable GWs, and also compatible with EWBG. It will be shown that for strong deflagrations, the fluid velocity in front of the wall saturates and even decreases with increasing wall velocity $v_w$. Since the walls become thinner at the same time, the baryon asymmetry is enhanced at larger wall velocities for these transitions, becoming {\em positively} correlated with a strong GW signal. {\color{black}Despite this positive correlation, we find that producing the observed baryon asymmetry together with a GW signal detectable in next generation observations is not possible, in contrast to previous estimates~\cite{vaskonen2017electroweak,Xie:2020wzn}. The difference comes from several factors working in the same direction. For example, we find larger wall velocities and thicknesses than Ref.\ \cite{vaskonen2017electroweak}, which suppress the baryon asymmetry. Moreover, our GW fits include a recently derived suppression factor due to shock reheating~\cite{Guo:2020grp,Hindmarsh:2020hop}, which leads to a much weaker GW signal for strong deflagrations.}\\

A further improvement in this work is to present an ultraviolet completion of the effective coupling that gives rise to the CP-violation needed for EWBG.  We introduce heavy vectorlike top partners which when integrated out induce a CP-violating coupling of the singlet scalar $s$ to top quarks,  giving the source term for EWBG.\footnote{Hints of the presence of such a particle in LHC data were recently presented in Ref.\ \cite{Waltenberger:2020ygp}.}\ \
Although the effective operator description of this term is quite adequate for quantitatively understanding EWBG \cite{deVries:2017ncy,Postma:2020toi},
its resolution in terms of underlying physics is necessary for quantifying how large its coefficient can be,  consistent with laboratory constraints.
We present the details in section \ref{sec:model}, including comprehensive collider limits on the top partners and the subsequent constraints on the effective theory.  The
finite-temperature effective potential of the theory is also outlined there, along with a discussion of cosmological constraints on the small explicit breaking of the $Z_2$ symmetry, that is necessary for EWBG.\\

The paper continues in Sect.\ \ref{sec:PTBN} with a brief description of our methodology for finding the high-temperature first-order phase transitions, and characterizing their strength. This is followed in Sect.\ \ref{sect:shape} by a detailed account of how the bubble wall speed
and shape are determined.  The techniques for
computing the baryon asymmetry and GW production are described in Sect.\ \ref{sec:sig}.  We present the results of a Monte Carlo exploration of the model parameter space with respect to these observables in 
Sect.\ \ref{sec:MC}, with emphasis on the interplay between successful EWBG and potentially observable GWs.  Conclusions are given in Sect.\ \ref{sec:conc}, followed by several appendices containing details about construction of the finite-temperature effective potential, solving junction conditions for the phase transition boundaries, and predicting GW production.

\section{$Z_2$-symmetric singlet model}
\label{sec:model}

 \par We study the $Z_2$-symmetric singlet scalar extension of the SM with a real singlet $s$  coupled to the Higgs doublet $H$. The scalar potential is
\begin{equation}\label{model}
\begin{aligned} V(H, s)= \mu_{h}^{2} H^{\dagger} H+\lambda_{h}\left(H^{\dagger} H\right)^{2} +\frac{\lambda_{hs}}{2}\left(H^{\dagger} H\right) s^{2}+\frac{\mu_{s}^{2}}{2} s^{2}+\frac{\lambda_{s}}{4} s^{4}.
\end{aligned}
\end{equation}
We work in unitary gauge, which consists of taking $H=h/\sqrt{2}$; the Goldstone bosons still contribute to the one-loop and thermal corrections, but they are set to zero in the tree-level potential. We assume $\mu_h^2<0$ and $\mu_s^2<0$, which implies that the potential has non-trivial minimums at $v\equiv h=\pm |\mu_h|/\sqrt{\lambda_h}\approx 246$ GeV, $s=0$ and $h=0$, $s=\pm |\mu_s|/\sqrt{\lambda_s}$. The scalar fields' mass in the vacuum can then be written in terms of the parameters of the potential as $m_h^2=-2\mu_h^2\approx (125\ \mathrm{GeV})^2$ and $m_s^2=-{\lambda_{hs}\mu_h^2}/({2\lambda_h})+\mu_s^2$. \\

The other relevant interaction of $s$
is a dimension-5 operator yielding an imaginary contribution to the top quark mass \cite{Cline:2012hg}:
\begin{equation}\label{top_quark_mass}
{\mathcal L}_{BG}=-\frac{y_{t}}{\sqrt{2}} h \overline{t}_{L}\left(1+i \frac{s}{\Lambda}\right) t_{R}+\mathrm{H.c.}
\end{equation}
This term will be ignored during the discussion on the phase transition; however it is essential for generating the baryon asymmetry, since it gives the CP-violating source term when $s$ temporarily gets a VEV in the bubble
walls of the electroweak phase transition.
In Eq.\ (\ref{top_quark_mass}) we have adopted a special limit of a more general model, in which the dimension-5 contribution is purely imaginary.
This can be understood as a consequence of imposing CP in the effective Lagrangian, with $s$ coupling like a pseudoscalar, $s\to -s$.  Hence it is consistent to omit terms odd in $s$ in the scalar potential
(\ref{model}), even though Eq.\ (\ref{top_quark_mass}) is odd in $s$.
The CP symmetry prevents a VEV from being generated for $s$ by loops.\\

The effective operator is generated by integrating out a heavy singlet vectorlike top quark partner $T$, whose mass term and couplings to the third generation quarks $q_L=(t_L,b_L)$, Higgs and singlet fields are
\be
   y_t \bar q_L H t_R + \eta_1 \bar q_L H T_R + 
   i\eta_2 \bar T_L s t_R + M\bar T_L T_R + {\rm H.c.}
\ee
including also the SM $q_L$-Higgs coupling.  This is invariant under
$CP$ if $s\to -s$.\footnote{The interaction term $i \eta_3 {\overline T}_L s T_R$ also respects CP for real $\eta_3$.
We neglect it to simplify our analysis.}\ \  
Integrating out $T$ leads to the effective operator in (\ref{top_quark_mass}) with scale
\be     
    \Lambda = {y_t M\over\eta_1\eta_2}\, .
\label{eq:Lambda}
\ee
We consider experimental constraints on the scale $\Lambda$ below. \\ 

In previous literature, thermal corrections were frequently approximated  by including just the first term of the high-temperature expansion of the thermal functions presented in the Appendix B. However, this approximation fails at temperatures below the mass of particles strongly coupled to the Higgs, as can happen in  models with a high degree of supercooling. Therefore, we employ the full one-loop thermal functions. This will be shown to have a large impact on the values of the tunneling action, and thus of the nucleation temperature. In addition to the tree-level potential and the thermal corrections, we also include the one-loop correction and the thermal mass Parwani resummation \cite{parwani1992resummation}. The complete effective potential then becomes
\begin{equation}
V_{\mathrm{eff}}=V_{\mathrm{tree}}+V_\mathrm{CW}+V_T+\delta V.
\end{equation}
The details  are presented in Appendix \ref{app:ep}. 

\subsection{Laboratory constraints}
\label{sub:lab}
It is important to determine how low the scale $\Lambda$ of the dimension-5 operator in Eq.\ (\ref{eq:Lambda}) can be, since it has a strong impact on the baryon asymmetry $\eta_b$; in the limit of large $\Lambda$, $\eta_b$ scales as $1/\Lambda$.  The relevant masses and couplings are constrained by direct searches for the top partner and precision electroweak studies.  Moreover the properties of the singlet $s$ are constrained by collider
searches. \\ 

\begin{figure}[t!]
	\centering
	\includegraphics[width=0.55\textwidth]{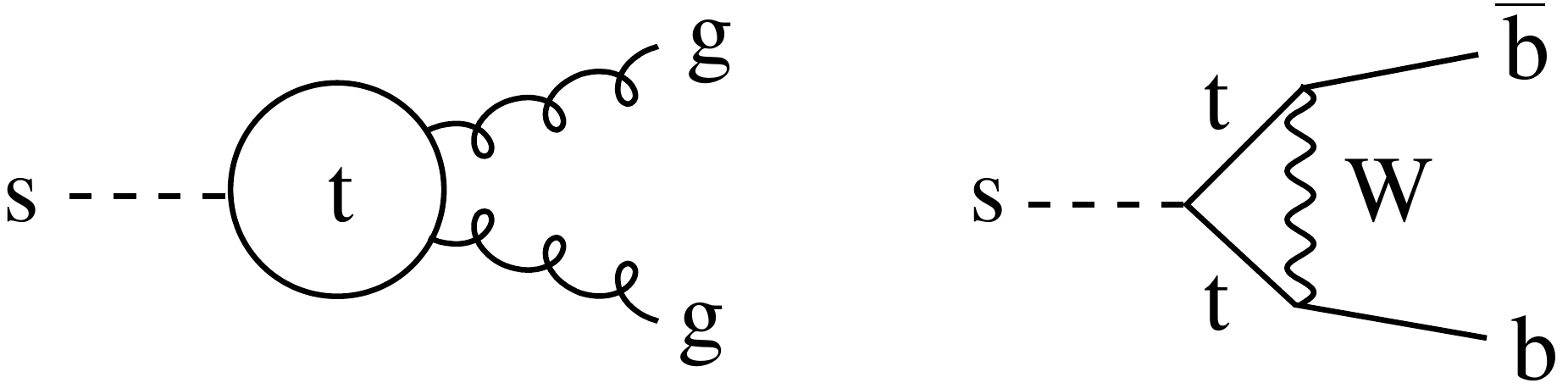}
	\caption{Feynman diagrams for decay of the singlet $s$.  The decay into gluons is by far the dominant channel.
	}
	\label{fig:loop}
\end{figure}

After electroweak symmetry breaking, a Dirac mass term  $(\bar t_L,\bar T_L)({m_t\atop 0}\,{\mu\atop M})\left({t_R\atop T_R}\right)$ is generated for $t,T$, with $m_t = y_t v/\sqrt{2}$ and $\mu = \eta_1 v/\sqrt{2}$ that is diagonalized by separate rotations
on $(t_R,T_R)$ and $(t_L,T_L)$, with mixing angles
\be
    \tan2\theta_L = 2 {M \mu\over M^2 - m_t^2 -\mu^2},\quad
    \tan2\theta_R = 2 {m_t \mu\over M^2 + \mu^2 - m_t^2}\,.
\ee
For example, we consider a benchmark point with $\eta_1=0.55$ and a physical $T$ mass $M_T = 800$ GeV, which correspond to $M=794\,$GeV and mixing angles $\theta_L=0.126$ and $\theta_R=0.027$.  The  relations between $y_t$ and the physical top mass differ from the SM ones by less than 1\%, which is allowed
by current LHC constraints \cite{Sirunyan:2018koj,Aad:2019mbh}.
For sufficiently large $\eta_2$, 
decays of $T$ to $h t/Z t/W b$ induced by mixing are highly subdominant to $T\to s t$, and searches for vector-like top partners that focus on the former channels are evaded. 
Near the Goldstone-equivalent limit (which should apply reasonably well for $M_T = 800$ GeV and relatively small $s$ masses, $m_s \sim 100$ GeV), the branching ratio for $T\to s t$ is
\begin{equation}
    B(T \rightarrow s t) \simeq \frac{\eta_2^2}{\eta_2^2 + 2 \eta_1^2}\,.
\end{equation}
We roughly estimate from Refs.\ \cite{Aaboud:2018pii,Sirunyan:2018omb} that for $M_T = 800$ GeV, vector-like quark searches that target SM final states are evaded provided $B(T \rightarrow s t) \gsim 90\%$, corresponding to $\eta \gsim 2.4$ for our benchmark point.  
Ref.\
\cite{Brooijmans:2020yij} (see Fig.\ 1 of contribution 5; 
also \cite{Cacciapaglia:2019zmj}) has reinterpreted collider bounds to constrain the parameter space $(m_s,M_T)$ for models in which $T \rightarrow s t$ dominates, finding that top partner masses above $\sim$ 750 GeV are allowed in the case
where $s$ decays 100\% into two gluons.  This is true in our model, where the dominant $s$ decays are induced by the loop diagrams shown in Fig.\ \ref{fig:loop}.  One can estimate that the
gluon final state dominates over that of $b$ quarks by a factor of
$(g_s^2 m_s/g_w^2 m_b)^2 \gtrsim 10^3$, and over decays into photons
by $(g_s/e)^4\sim 300$.
Precision electroweak data constrain the additional contributions to the
oblique parameters, especially $T$, which is corrected by \cite{Dawson:2012di}
\be
\Delta T = T_{\rm sm} s_L^2 \left( -(1 + c_L^2) + s_L^2\, r + 2 c_L^2\, {r\over r - 1}\ln r\right)
\lesssim 0.1\,,
\ee
where $T_{\rm sm} = 1.19$ is the SM value, $c_L=\cos\theta_L$, $s_L = \sin\theta_L$, and $r = (M_T/m_t)^2$; the upper limit is from section 10 of \cite{Tanabashi:2018oca}. The benchmark point chosen above almost saturates this constraint, giving $\Delta T \simeq 0.09$.\\

\begin{figure}[t!]
	\centering
\centerline{\includegraphics[width=0.5\textwidth]{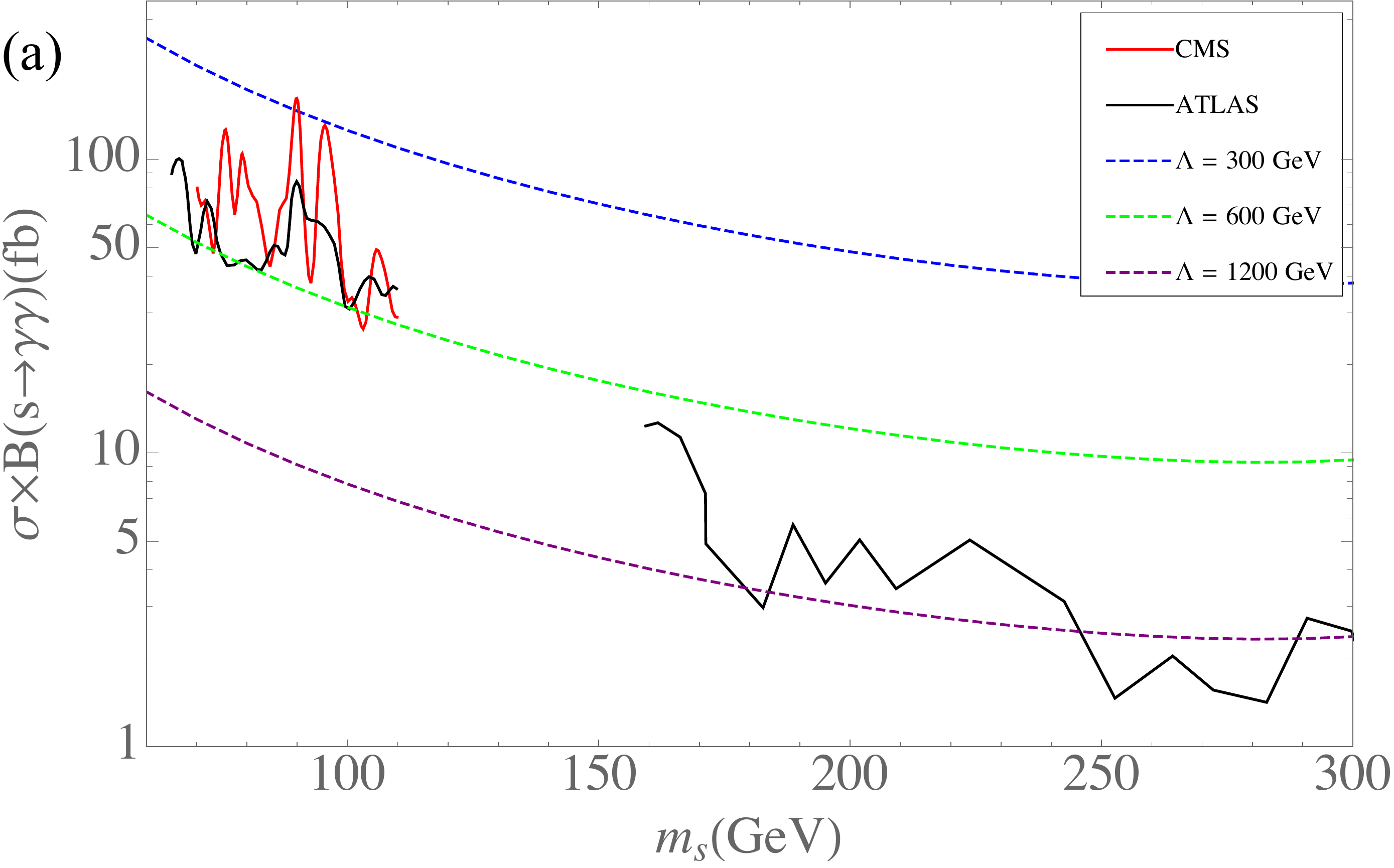}
\quad\quad
\includegraphics[width=0.5\textwidth]{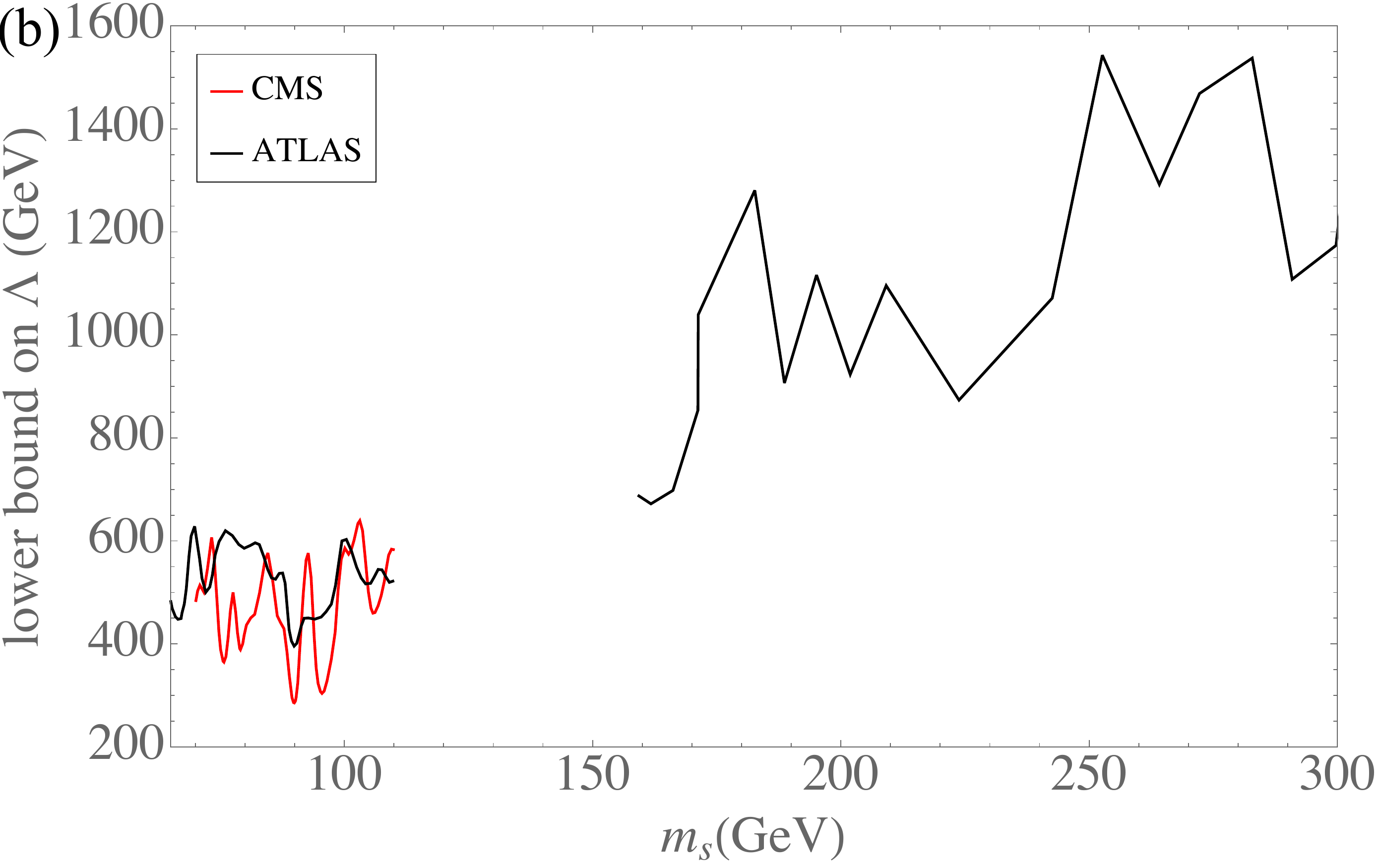}}
\vskip-0.5cm
\leftline{(a) \hspace{8.5cm} (b)}
		\caption{Left (a): experimental limits from ATLAS \cite{ATLAS:2020tws,ATLAS:2018xad} and CMS \cite{Sirunyan:2018aui} for resonant production of $s$ by $gg$ fusion followed by decays into photons (solid lines), versus predictions at different values of of $\Lambda$.  Right (b): corresponding lower bounds on $\Lambda$. 	}
	\label{fig:diphoton}
\end{figure}

There are also direct searches for resonant production of the singlet,
by gluon-gluon fusion.  
The coupling of $s$ to $t$ in the mass eigenstate basis is $y_{st} = \eta_2\cos\theta_R\sin\theta_L\sim \eta_2\theta_L$, while that to $T$ is
$y_{sT} = -\eta_2\cos\theta_L\sin\theta_R\sim -\eta_2\theta_R$.  
The squared matrix element for the decays $s\to gg$ is \cite{Craig:2015lra}
\be
    |{\cal M}|^2 = \left(\alpha_s\over \pi\right)^2 m_s^4\left|
    \sum_{i=t,T} {y_{si}\over m_i}\,\tau_i \left[\sin^{-1}\left( \tau_i^{-1/2}\right)\right]^2\right|^2, 
\ee
where $\tau_i = 4 m_i^2/m_s^2$.  The parton-level production cross section for $gg\to s$ is 
    $\hat\sigma = {\pi}|{\cal M}|^2\delta(\hat s - m_s^2)/(256\, \hat s)$
where the 256 comes from averaging over gluon colors and spins.  Integrating this over the gluon PDFs gives the hadron-level
cross section
\be
    \sigma(pp\to s) = {\pi\over 256\, m_s^4}|{\cal M}|^2 {\cal L}_g \equiv
    {\pi\over 256\, m_s^4}|{\cal M}|^2 \int_{m_s^2/s}^1
    {dx\over x}  [x f_g](x)[x f_g](m_s^2/sx) 
\ee
in which dependence on $m_s$ drops out except in the parton luminosity factor ${\cal L}_g$.
This production is probed via decays $s\to\gamma\gamma$, whose branching ratio is approximately $B(s\to\gamma\gamma) = (8/9)\alpha^2/\alpha_s^2$  \cite{Craig:2015lra}.  For the dominant $s\to gg$ decay into gluons, in principle LHC dijet resonance searches could be constraining, but these exist only for $m_s\gtrsim 500\,$GeV which is beyond the range of interest for the present study.
To a good approximation, $\sigma(pp\rightarrow s)$ is determined by $m_s$ and $\Lambda$.  
In Fig.\ \ref{fig:diphoton}(a) we show limits from ATLAS \cite{ATLAS:2020tws,ATLAS:2018xad} and CMS \cite{Sirunyan:2018aui} on $\sigma B(s\to\gamma\gamma)$ as a function of $m_s$, along with the predictions for various $\Lambda$, and in Fig.\ \ref{fig:diphoton}(b) we show the associated lower bounds on $\Lambda$. 
In the low-mass region ($65$ GeV $<m_s<110$ GeV), lower bounds on $\Lambda$ range roughly from 400 GeV to 650 GeV; in the intermediate-mass region ($110$ GeV $<m_s<160$ GeV), $\Lambda$ is not yet constrained by diphoton resonance searches, and for much of the high-mass region ($m_s> 160$ GeV), $\Lambda$ is bounded to be above 1 TeV. For our subsequent scans of parameter space, we adopt a fixed reference value for $\Lambda$,
\begin{equation}
    \Lambda_\text{ref} = 540 \text { GeV},
\end{equation}
which is large enough to be consistent with much of the low-$m_s$ region.  Because $\Lambda_\text{ref}$ is well below the lower-bounds on $\Lambda$ in the high-mass region,  we confine our scans to $m_s < 160$ GeV for consistency.\footnote{Although we do not pursue this point here,  lower values of $\Lambda$ are consistent with $m_s > 160$ GeV if  $B(s \rightarrow \gamma\gamma)$ is suppressed, for example by a dominant invisible decay channel; LHC constraints on $t{\overline t}$ plus missing energy \cite{Aad:2021hjy,Sirunyan:2017leh}
are in that case evaded for  $M_T \gsim 1350$ GeV.}\\

\begin{figure}[t!]
	\centering
\centerline{\includegraphics[width=0.4\textwidth]{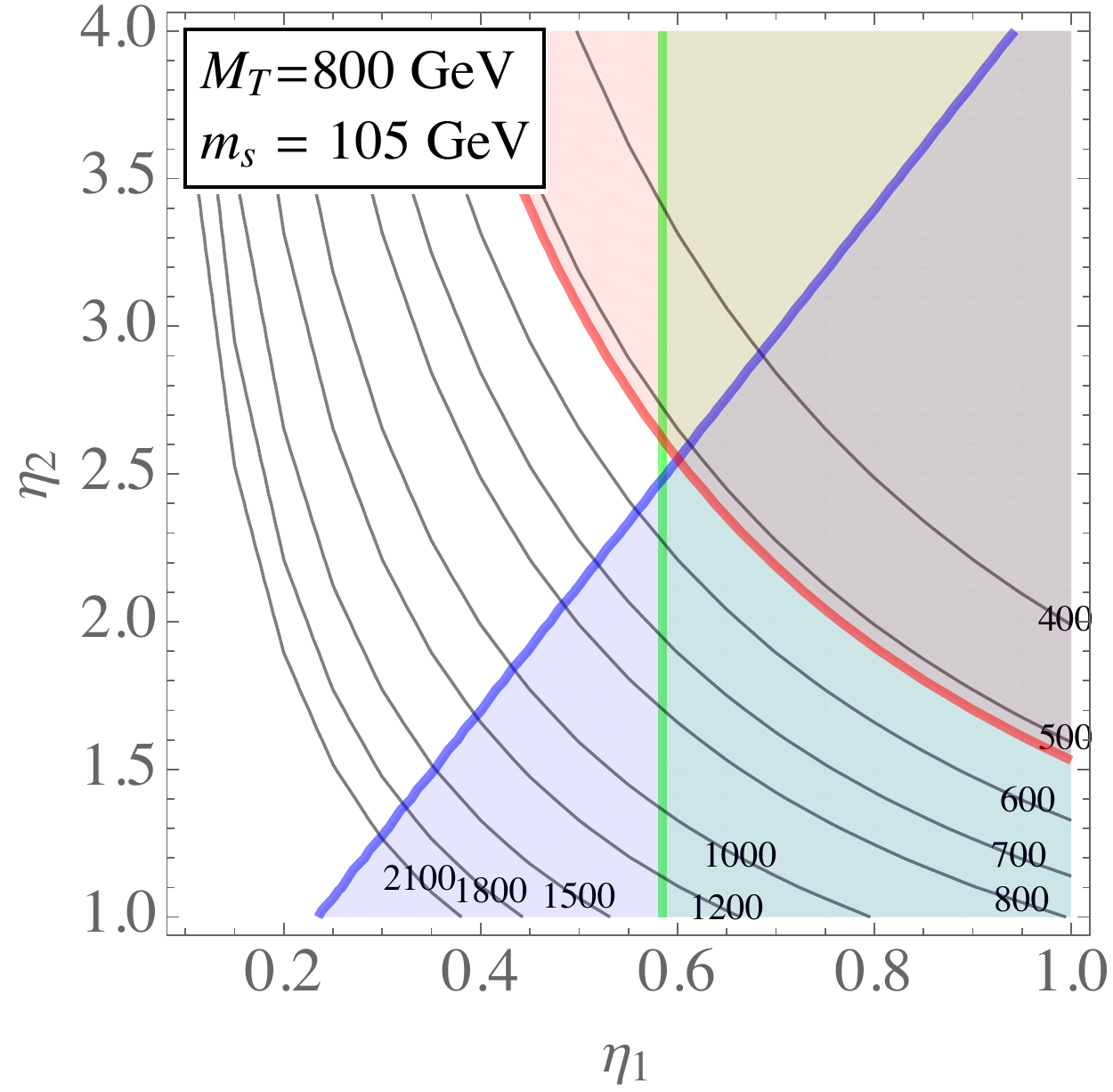}
\quad\quad
\includegraphics[width=0.4\textwidth]{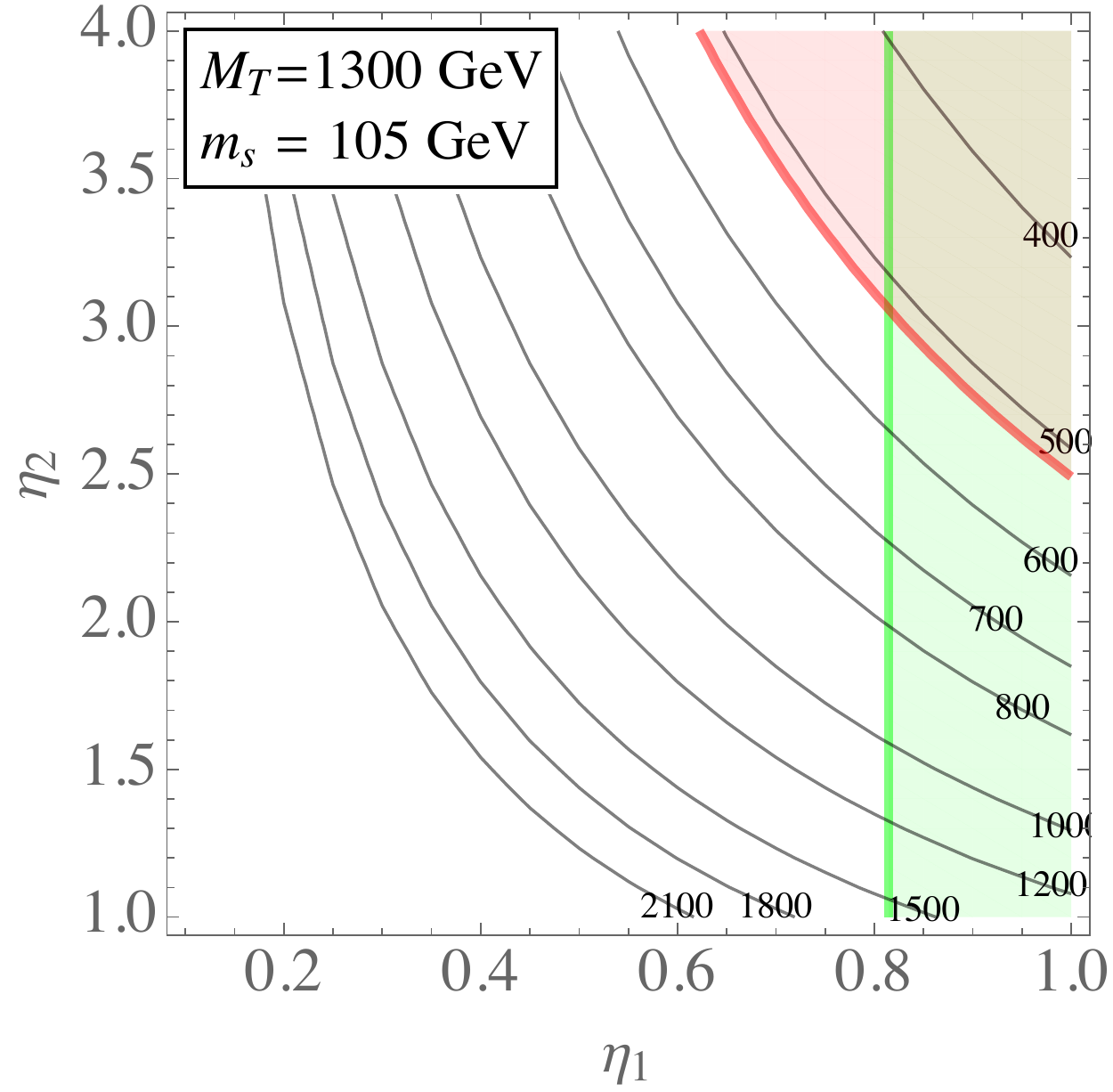}}
\vskip-0.5cm
	\caption{For selected $T$ and $s$ masses, constraints on $\eta_1$ and $\eta_2$ from precision electroweak data (green), diphoton resonance searches	\cite{Sirunyan:2018aui,ATLAS:2018xad}(red), and searches for vector-like quarks \cite{Aaboud:2018pii} (blue), along with contours of $\Lambda$ in GeV.  The allowed region is unshaded. 	}
	\label{fig:etaspace}
\end{figure}

The constraints from precision electroweak data, diphoton resonance searches, and vector-like quark searches  are shown in the $\eta_1$-$\eta_2$ plane in Fig.~\ref{fig:etaspace}, for $M_T = 800$ GeV, where we approximate the $T$ search constraints by the requirement $B(T\rightarrow st) > 0.9$,  and for $M_T = 1300$ GeV, heavy enough to evade $T$ searches for any $B(T\rightarrow st)$.  For the chosen $m_s$, it is apparent that the reference value $\Lambda = 540$ GeV is attainable for $\eta_2 \gsim 2.5$ for $M_T = 800$ GeV and $\eta_2 \gsim 3$ for $M_T = 1300$ GeV. For slightly heavier $s$ in the window $110$ GeV $< m_s < 160$ GeV, diphoton resonance searches are evaded and the red contours disappear. In this case even lower values of $\Lambda$ are allowed provided one is willing to consider larger values of $\eta_2$. Since the baryon asymmetry $\eta_b$ scales roughly as $1/\Lambda$, it is straightforward to reinterpret our final results for larger (or smaller) $\Lambda$.  From the results of
Section \ref{sec:MC} one can infer that a significant fraction of models remain viable for baryogenesis for $\Lambda = 2 \Lambda_\text{ref}$ (or for even larger $\Lambda$), a scale consistent with more modest couplings, $\eta_2 \sim 1.5$.\\

Allowing for very large values of $\eta_2$ could invalidate the effective theory above
the heavy top partner threshold $M$ at scales only slightly larger than $M$, which would require us to specify additional new physics in order to have a complete description.  There are two principal challenges arising from the running of the couplings,
\bear
    {d\eta_2\over d\ln\mu} &\cong& {\eta_2^3\over 4\pi^2}\\
    {d\lambda_s\over d\ln\mu} &\cong& {9\,\lambda_s^2\over 8\pi^2} - {3\,\eta_2^4\over 2\pi^2}+\frac{\lambda_s \eta_2^2}{2 \pi^2}
\eear    
where $\mu$ denotes the renormalization scale.  The
most serious problem is that for large values of $\eta_2$, the self-coupling $\lambda_s$ is quickly driven to zero, and the scalar potential becomes unstable.
The second is that $\eta_2$ reaches a Landau pole at somewhat higher scales.  The first problem could be ameliorated by coupling additional scalars to $s$, without impacting our results for 
EWBG or GWs.  For this reason, we do not limit the scope of our investigation based on the running of $\lambda_s$.
Regarding the second problem, we note that even for $\eta_2=3$, the Landau pole is nearly an order of magnitude above $M$, which we consider to be an acceptably large range of validity for the effective theory.

\subsection{Explicit breaking of $Z_2$ symmetry}

Since we are considering a scenario where the $Z_2$ symmetry $s\to -s$ is spontaneously broken during the early universe and restored at the EWPT, 
domain walls form before the EWPT, and the universe will consist of domains with random signs of the $s$ condensate.  The source term for EWBG that arises from Eq.\ (\ref{top_quark_mass}) is linear in $s$, resulting in baryon asymmetries of opposite signs, that could average to zero after completion of the EWPT. To avoid this outcome, the $Z_2$ symmetry should be explicitly broken, by potential terms
\be
V_b = \mu_b s (h^2 - v^2) + \mu_b's^3
\ee
with small coefficients $\mu_b$, $\mu_b'$. We have used the freedom of shifting $s$ by a constant to remove a possible tadpole of $s$ at the true vacuum $(h,s)=(v,0)$. \\

The presence of the biasing potential $V_b$ can prevent the baryon washout in several ways. First, if the transition to the broken-$s$ phase is of second order, even a small tilt can suffice to make the lower-energy vacuum dominate.  Second, in a first order transition, symmetry breaking terms can bias the bubble nucleation rates to prefer the lower-energy vacuum. Indeed, the number of bubbles nucleated during the transition is $n \sim \int_{t_c}^{t_*} {\rm d}t\,\Gamma(t)$, where $t_*$ is the time when transition completes, and $\Gamma(t) \sim \exp(-S_3/T)$. Writing the action as $S_{3\pm}=\bar S_3 \mp \delta S$ in the two respective vacua, the relative number density of bubbles in each phase at the end of the transition becomes $n_+/n_- \approx \exp(2\,\delta S_*/T_*)$. In general~\cite{Enqvist:1991xw} $S_3 \propto E$, where $E$ is the coefficient of the cubic term in the potential. Using this scaling we may write $\delta S_* = (\delta E/E_0)\bar  S^*_3$, where typically $S^*_3/T_*\approx 100$.  In our model $E_0 \approx (3\lambda_s)^{3/2}T/12\pi$, so taking $V_b = \mu_{b'}s^3$, corresponding to $\delta E = \mu_{b'}$, and $T_* \approx 100\,$GeV, the condition for single-phase vacuum dominance becomes $\mu_{b'} \gsim 0.1\,\lambda_s^{3/2}$\,GeV. Barring very large $\lambda_s$, this condition is easily met with no limitations on our analysis.\\

Even if a domain wall network forms, the higher-energy domains will collapse due to pressure gradients, and we should ensure that this process completes before the EWPT. The collapse starts with the acceleration of a wall at relative position $R$ according to $\ddot R = -\Delta V/\tau$, where $\tau \sim \sqrt{\lambda_s} w^3$ is the surface tension (distinct from the  tension $\sigma$ used above in the nucleation estimate), $\Delta V \sim V_b(0,w) \sim \mu_b'w^3$ is the difference in the vacuum energies, and $w\sim \mu_s/\sqrt{\lambda_s}$ is the singlet VEV. Using $H =1/2t$ and $T\approx 100$\,GeV, one finds that walls reach light speed in time 
\be
{\delta t\over t} = {\tau H\over \delta V} \sim 10^{-5}\sqrt{\lambda_s}\left({\rm eV}\over\mu_{b'}\right)\,,
\label{dtot}
\ee
which is practically instantaneous on the timescales of interest, for reasonable values of $\mu_{b'}$.
We note that global symmetries like $Z_2$ are expected to be broken by quantum gravity effects, so that it could be reasonable to anticipate $\mu'_b \sim v^2/M_p \sim 0.1\,$eV, which is large enough from the perspective of Eq.\ (\ref{dtot}).\\

The higher energy domains subsequently collapse at the speed of light, since there is no appreciable friction. The time required for this process to complete is determined by 
    $R_* =2 a(t_1) \int_{t_1}^{t_2} {dt}/{a(t)}$,
where $R_*$ is the comoving size of the domain wall separation.  By the Kibble mechanism one expects that $R_* = AH_*^{-1}$ with $A\lesssim 1$, leading to the ratio of domain wall collapse to formation times $t_2/t_1 = ( 1 + A/2)^2$. The temperature interval corresponding to this time interval is ${\Delta T/T} \approx A$, assuming that the growth phase also proceeded at the speed of light.\\

The temperature of the first phase transition, $T_1$ can be estimated as that when $\partial^2 V/\partial s^2$ becomes negative. In the approximation of neglecting $V_b$, and keeping only leading terms in the high-$T$ expansion, one finds $T_1^2 - T_c^2 \sim \lambda_h w_c^2/c_s$
where $T_c$ is the critical temperature of the EWPT, and $c_s = \left(3\lambda_s 
+ 2\lambda_{hs} \right)/12$. Thus the temperature difference between transitions is of order
$\Delta T_{1c} \sim \lambda_h w^2 / (c_s T_c)$. Requiring that $\Delta T_{1c}/T_c > A$ then gives
\be	
A < \frac{12\lambda_h}{3\lambda_s + 2\lambda_{hs}}\,\frac{w_c^2}{T_c^2}\sim O(1)\,.
\ee
Given that $A \sim (T_*/S^*_3)(\Delta T/T)_* \sim 10^{-2}$-$10^{-4}$~\cite{Moore:1995si}, this is a very weak constraint. We conclude that it is easy to avoid cosmological problems associated with the domain walls by small symmetry breaking terms, that do not affect the rest of our analysis.

\section{Phase Transition and Bubble Nucleation}
\label{sec:PTBN}
In the examples of interest for this work, 
the phase transition in the $Z_2$-symmetric singlet model
proceeds in two steps: starting from the high-temperature
global minimum $h=s=0$, a transition first occurs to 
nonzero $s$, while the Higgs field remains at $h=0$.  
This is followed by the EWPT, in which $s$ returns to zero and $h$ develops its VEV.  The $h^2s^2$ interaction provides the potential barrier to make this a first order transition.\\

As usual, the first order transition occurs at the bubble nucleation temperature $T_n$, which is below the critical temperature $T_c$, where the two potential minima become degenerate,
\begin{equation}
\left.	V_{\mathrm{eff}}(h, s,T_c)\right|_{h=0,\atop s=w_c}=\left.V_{\mathrm{eff}}(h,s,T_c)\right|_{h=v_c,\atop s=0}
\end{equation}
Bubble nucleation occurs when the vacuum decay rate per unit volume $\Gamma_d$ becomes comparable to $H^4$, the Hubble rate per Hubble volume.
The decay rate is \cite{Linde:1980tt}
\begin{equation}
\Gamma_d \cong T^{4}\left(\frac{S_{3}}{2 \pi T}\right)^{3 / 2} \exp \left(-\frac{S_{3}}{T}\right)\,,
\end{equation}
where $S_3$ is the O(3) symmetric action,
\begin{equation}
S_{3}=4 \pi \int r^{2} d r\left(\frac{1}{2}\left(\frac{d h}{d r}\right)^{2}+\frac{1}{2}\left(\frac{d s}{d r}\right)^{2}+V_{\rm eff}\right )\,.
\end{equation}
The precise criterion that we use for nucleation is 
\begin{equation}\label{nucleation}
\exp \left(-S_{3} / T_{n}\right)=\frac{3}{4 \pi}\left(\frac{H\left(T_{n}\right)}{T_{n}}\right)^{4}\left(\frac{2 \pi T_{n}}{S_{3}}\right)^{3 / 2}\,,
\end{equation}
which is satisfied when $S_3/T_n\cong 140$ \cite{Quiros:1999jp}.
We used the package CosmoTransitions \cite{wainwright2012cosmotransitions} to calculate $S_3$. The action obtained with the full potential can differ significantly from the commonly used thin wall approximation \cite{cline2017electroweak,coleman1977fate} or the approximation of evaluating it along the minimal integration path for the potential \cite{vaskonen2017electroweak}. We compare the predictions for nucleation of these approximations to the full one-loop result, for several exemplary models, in Table \ref{table1}. The approximate methods tend to underestimate the action, giving a higher nucleation temperature; hence we use the values derived from the full one-loop action in the following.\\

\begin{table}[ht]
\label{table1}
	\centering
	\begin{tabular}{|c|c|c|c|c|c|c|c|}
		\hline
		\multirow{2}{*}{$\lambda_{hs}$} & \multirow{2}{*}{$m_s\ (\mathrm{GeV})$} & \multicolumn{3}{c|}{$S_3/T\vert_{T=100\ \mathrm{GeV}}$} & \multicolumn{3}{c|}{$T_n\ (\mathrm{GeV})$}\\
		\cline{3-8}
		 & & Thin wall & MPP & 1-loop & Thin wall & MPP & 1-loop \\
		\hline
		1 & 120 & 234 & 277 & 427 & 93.5 & 92.6 & 89.8\\
		\hline
		1.7 & 200 & 68.7 & 101 & 151 & 115.6 & 109.8 & 100.1\\
		\hline
		3.2 & 300 & 37.9 & 36.8 & 54.3 & 134.3 & 133.8 & 121.6\\
		\hline
	\end{tabular}
	\caption{\small{Examples of the dimensionless tunneling action $S_3/T$, evaluated at $T=100$\,GeV, and ensuing nucleation temperatures, computed within the thin wall and minimal potential path (MPP) approximations, compared with the value obtained using the resummed one-loop potential. In there example, $\lambda_s=1$ and $\Lambda=540\ \mathrm{GeV}$.}}
\end{table}

There are two complementary parameters for characterizing the strength of the first order transition.  One is the
ratio of the Higgs VEV to the temperature at the time of
nucleation, $v_n/T_n$, which is especially relevant for
EWBG, as we will discuss in Sect.\ \ref{sec:EWBG}.   The other, which is more important for GW production, is the ratio of released vacuum energy density to the radiation energy density \cite{kamionkowski1994gravitational,espinosa2010energy}:
\begin{equation}
\alpha=\frac{1}{\rho_{\gamma}}\left(\Delta V-\frac{T_n}{4} \Delta \frac{d V}{d T}\right),
\label{alphadef}
\end{equation}
where $\rho_\gamma=g_*\pi^2T_n^4/30$, $g_*$ is the effective number of degrees of freedom in the plasma (we use $g_*=106.75$) and $\Delta$ denotes the difference between the unbroken and broken phase. $\alpha$ quantifies the amount of supercooling that occurs prior to nucleation, which determines how much free energy is available for the production of GWs.\\

\section{Wall velocity and shape}
\label{sect:shape}

The derivation of the wall velocity and field profiles is
a technically demanding problem \cite{Moore:1995si}, that was first addressed
in the context of Higgs plus singlet models 
in Refs.\ \cite{Huber:2013kj,Konstandin:2014zta,Kozaczuk:2015owa}, in various approximations.
One must solve the equations of motion (EOM) for the scalar sector coupled to a perfect fluid,
\bea
E_h(z) &\equiv -h''(z)+\frac{dV_\mathrm{eff}(h,s;T_+)}{dh}+\sum\limits_i N_i \frac{d m_i^2}{dh}\int\frac{d^3 p}{(2\pi)^3 2E}\,\delta f_i(\Vec{p},z) = 0,\\
E_s(z) &\equiv -s''(z)+\frac{dV_\mathrm{eff}(h,s;T_+)}{ds}
+\sum\limits_i N_i \frac{d m_i^2}{ds}\int\frac{d^3 p}{(2\pi)^3 2E}\,\delta f_i(\Vec{p},z) = 0,
\label{eq:EOMs}
\eea
where $z$ is the direction normal to the wall, that is to a good approximation planar by the time it has reached its terminal velocity. We use a sign convention where the wall is moving to the left, so that $z>0$ corresponds to the broken phase.  The sum is over all the relevant species coupled to $h$ or $s$ in the plasma, with $N_i$ and $m_i$  respectively denoting the number of degrees of freedom and the field-dependent mass of the corresponding species, and $\delta f_i$ the deviation from equilibrium of its distribution function. All the temperature-dependent quantities appearing in these equations are evaluated at $T_+$, which is the plasma's temperature just in front of the wall. We calculate $T_+$ in Appendix \ref{app:fluidEq} using the method described in Ref.\ \cite{espinosa2010energy}, and $\delta f_i$ will be computed in Sect.\ \ref{sec:BE}.\\

The terms in Eqs.\ (\ref{eq:EOMs}) with $\delta f_i$ represent the friction\footnote{The term ``friction'' is strictly speaking not correct, but we adopt this commonly used terminology. More accurately, the last terms in~\eqref{eq:EOMs} represent the additional pressure created by the out-of-equilibrium perturbations, which modify the effective action in the same way as the usual thermal excitations.} of the plasma on the wall, that leads to a terminal wall speed $v_w<1$, unless the friction is too small and the wall runs away to speeds close to that of light.  Following previous work, we take the dominant sources of friction to be from the top quark ($i=t$) and electroweak gauge bosons ($i=W$), neglecting the contributions to friction from the Higgs itself and from the singlet.  This approximation is bolstered by the smaller number of degrees of freedom $N_h=N_s=1$ compared to $N_t = 12$ and $N_W = 9$, as well as the smallness of the  Higgs self-coupling $\lambda_h$ and the not-too-large values of the cross-coupling $\lambda_{hs}$ that will be favored in the subsequent analysis.  Then the friction term for the $s$ equation of motion vanishes, since $s$ couples only to itself and to the Higgs, apart from its suppressed dimension-5 coupling
to $t$.  This allows for some simplification in the following procedure.\\

In 
Ref.\ \cite{Friedlander:2020tnq}, a similar study of the present model was done, where no {\it a priori} restriction of the wall shape  was assumed, but it was found that the actual shapes conform to a very good approximation to the tanh profiles
\bea
h(z) &= \frac{h_0}{2}[1+\tanh(z/L_h)], \\
s(z) &= \frac{s_0}{2}[1-\tanh(z/L_s+\delta)],
\label{shapedef}
\eea
where $h_0$ and $s_0$ are respectively the vacuum expectation values (VEV) of the $h$ and $s$ fields in the broken and unbroken phases.
Hence we adopt the ansatz (\ref{shapedef}),
which allows the singlet and Higgs wall profiles to have different widths, and to be offset from each other by a distance $L_s\delta$. The $s$ field's VEV is taken to be the usual one evaluated at $T_+$, which solves the equation ${dV_\mathrm{eff}(0,s;T_+)}/{ds}\big\vert_{s=s_0}=0$. 
The situation is more complicated for the $h$ field, for which the Higgs VEV should be evaluated at $T_-$, the plasma's temperature behind the wall. Since we are fixing a constant temperature $T_+$ in the potential, the change in the effective action due to the shift in the background temperature must be accounted for by the perturbation in the broken phase. As a consequence we are choosing $h_0$ so that it solves the equation
\be \label{eq:adjusted_vev}
\left.\left(\frac{dV_\mathrm{eff}(h,0;T_+)}{dh} + \sum\limits_i N_i \frac{d m_i^2}{dh}\int\frac{d^3 p}{(2\pi)^3 2E}\,\delta f_i(\Vec{p},z) \right)\right\vert_{h=h_0, z\to\infty} = 0\,.
\ee
This choice guarantees that the Higgs EOM is satisfied far behind the wall. We will estimate the uncertainty of our results due to this approximation in Sect.\ \ref{sec:theoretical_uncertaineties}.\\

\begin{figure}[t]
	\centering
	\includegraphics[width=0.45\textwidth]{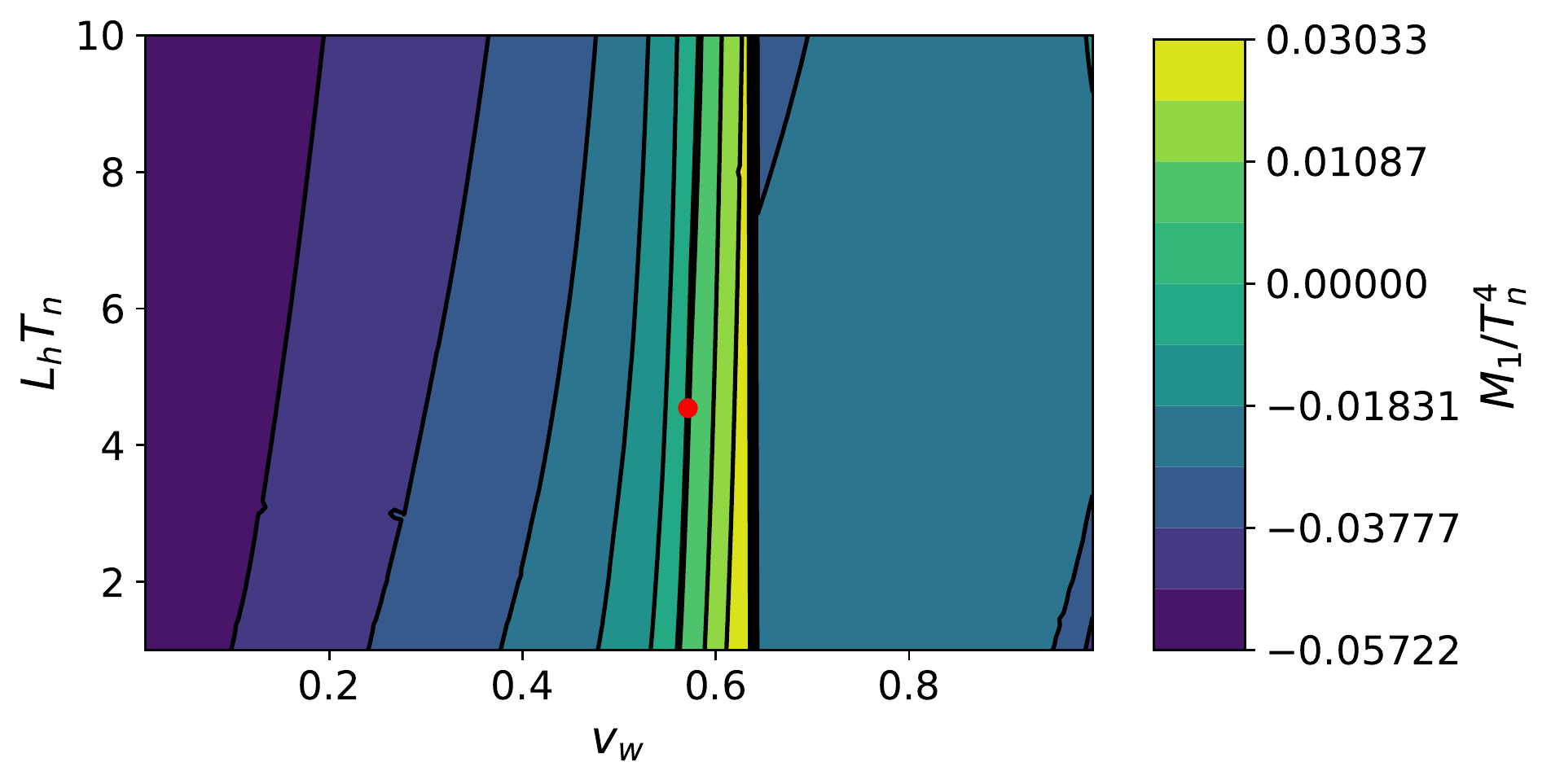}\hspace{5mm}\includegraphics[width=0.45\textwidth]{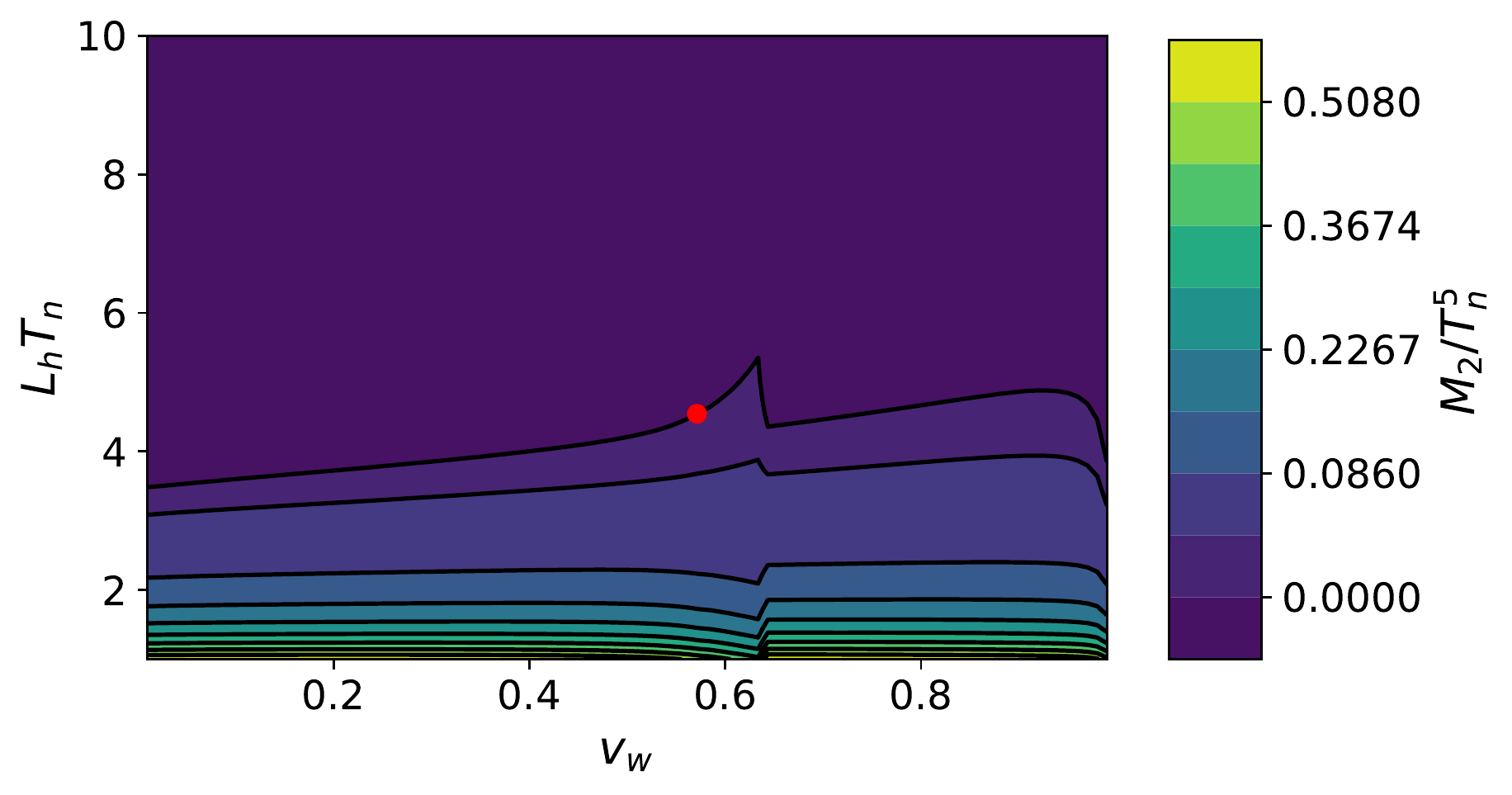}
	\centerline{ (a) \qquad\qquad\qquad\qquad\qquad\qquad\qquad\qquad\qquad\qquad\qquad\qquad (b) \qquad}
	\caption{\small{Moments of the Higgs EOM (a) $M_1$ and (b) $M_2$ as a function of the wall velocity $v_w$ and the Higgs wall width $L_h$ for a model with parameters $\lambda_{hs}=1$, $\lambda_s=1$ and $m_s=130\ \mathrm{GeV}$. The red dot is the solution of Eqs.\ (\ref{eq:moment1},\ref{eq:moment2}). As expected, $M_1$ is roughly independent of $L_h$ while $M_2$ depend mainly on $L_h$. The moments are discontinuous at $v_w\approx 0.63$ because this corresponds (for this specific model) to the boundary between hybrid and detonation walls, where $v_+$ and $T_+$ are discontinuous.}}
	\label{fig:contours}
\end{figure}

To approximately solve the Higgs EOM, one can define two independent moments $M_{1,2}$ of $E_h(z)$, and assume that they both vanish at the optimal values of  $v_w$ and $L_h$. A convenient choice is \cite{Konstandin:2014zta}
\bear 
\label{eq:moment1}
M_1 &\equiv& \int dz\, E_h(z)\, h'(z) = 0, \\
\label{eq:moment2}
M_2 &\equiv& \int dz\, E_h(z) [2h(z)-h_0]\, h'(z) = 0.
\eear
These also have intuitive physical interpretations that naturally distinguish them as good predictors of the
wall speed and thickness, respectively.
$M_1$ is a measure of the net pressure on the wall, so
that Eq.\ (\ref{eq:moment1}) can be interpreted as 
the requirement that a stationary wall 
should have a vanishing total pressure; nonvanishing
$M_1$ would cause it to accelerate.
Therefore one expects that Eq.\ (\ref{eq:moment1}) 
principally determines the wall speed $v_w$, while 
depending only weakly on the thickness $L_h$. With our sign convention, $M_1$ can be interpreted as the pressure in front of the wall minus the pressure behind it, so that $M_1>0$ corresponds to a net force slowing down the wall. On the other hand, 
$M_2$ is a measure of the pressure gradient in the wall. If nonvanishing, it would lead to compression or stretching of the wall, causing $L_h$ to change.
Hence Eq.\ (\ref{eq:moment2}) mainly determines 
$L_h$, and depends only weakly on $v_w$.
The two equations are approximately decoupled,
facilitating their numerical solution.  This is illustrated in Fig.\ \ref{fig:contours}, which shows the
dependence of $M_1$ and $M_2$ on $v_w$ and $L_h$.\\

We chose a different approach to determine the singlet wall parameters $L_s$ and $\delta$. Instead of solving moment equations analogous to (\ref{eq:moment1},\ref{eq:moment2}), one can determine
their values by minimizing the $s$ field action
\bea
S(L_s,\delta) &= \int dz\, \left\lbrace\frac{1}{2}(s')^2 + \left[V_{\mathrm{eff}}(h,s,T_+) - V_{\mathrm{eff}}(h,s^*,T_+)\right]\right\rbrace \\ 
&= \frac{s_0^2}{6L_s} + \int dz\, \left[V_{\mathrm{eff}}(h,s,T_+) - V_{\mathrm{eff}}(h,s^*,T_+)\right],
\eea
with respect to $L_s$ and $\delta$.  Here
 $s^*$ is a field configuration with arbitrary fixed parameters $L_s^*$ and $\delta^*$, that we choose to be $L_s^*=L_h$ and $\delta^*=0$.  The second term is just a constant, but it allows for the convergence of the integral by canceling the contributions of 
 $V_{\rm eff}$ at 
 $z\to\pm\infty$.
 This method has the advantage that it does not depend on any arbitrary choice of moments, and it is more efficient to numerically minimize the function of two variables than to solve the system of equations
 for the moments of the EOMs.\\

\subsection{Transport equations for fluid perturbations}
\label{sec:BE}
The final step toward the complete determination of the velocity and the shape of the wall is to compute the distribution functions' deviations from equilibrium $\delta f_i$, by solving the Boltzmann equation for each relevant species in the plasma. The method of approximating the full Boltzmann equation by a truncated 
set of coupled fluid equations was originally carried out
in Ref.\ \cite{Moore:1995si}, for the regime of slowly-moving walls (see also Ref.\ \cite{Konstandin:2014zta}).  This approach was recently 
improved in Ref.\ \cite{Laurent:2020gpg} in order to be
able to treat wall speeds close to or exceeding the speed of sound consistently.  We briefly summarize the formalism, which we use in the present study.\\

The out-of-equilibrium distribution function can be parametrized in the wall frame as
\be
f=\frac{1}{\exp[\beta\gamma(E-v_+ p_z)(1-\delta\tau) - \mu]\pm 1}+\delta f_u,
\ee
where $\beta=1/T_+$ and the $\pm$ is $+$ for fermions and $-$ for bosons. $\delta\tau$ and $\mu$ are the dimensionless temperature and chemical potential perturbations from equilibrium, and $\delta f_u$ is a velocity perturbation whose form is unspecified, but is constrained by $\int d^3p\, \delta f_u = 0$. By assuming that the perturbations are small, one can expand $f$ to linear order in $\mu$, $\delta\tau$ and the velocity perturbation $\delta f_u$ to obtain
\be \label{eq:perturbation}
\delta f \approx \delta f_u - f'[\mu+\beta\gamma\,\delta\tau(E-v_+ p_z)],
\ee
with 
\be
f'=\left. \frac{d}{dX}\frac{1}{e^X \pm 1}\right\vert_{X=\beta\gamma(E-v_+ p_z)}.
\ee 

To simplify the problem, one models the plasma as being made of three different species: the top quark, the $W$ 
bosons (shorthand for $W^\pm$ and $Z$) and a background fluid, which includes all the remaining degrees of freedom. It is convenient to write the velocity perturbation as $u\propto\int d^3p\, ({p_z}/{E})\,\delta\! f_u$ when constructing the moments of the linearized Boltzmann equation.
 By taking three such moments, using the weighting factors $1$, $E$ and $p_z/E$, the perturbations are determined by transport equations
\bear \label{eq:boltzmann1}
Aq'+\Gamma q&=&S,\\
\label{eq:boltzmann2}
q_\mathrm{bg}'&=&-\Tilde{A}_\mathrm{bg}^{-1}(\Gamma_{\mathrm{bg},t}q_t+\Gamma_{\mathrm{bg},\W}q_\W),
\eear
where prime denotes $d/dz$, $q_i=(\mu_i,\delta\tau_i,u_i)^\intercal$, $q = (q_\W^\intercal,q_t^\intercal)^\intercal$, the $\Gamma$ matrices are collision terms, and $S$ is the source term, whose definitions, as well as those of the the matrices $A$, $\Gamma$, $\Tilde{A}_\mathrm{bg}^{-1}$, $\Gamma_{\mathrm{bg},t}$, $\Gamma_{\mathrm{bg},\W}$,
can be found in Ref.\ \cite{Laurent:2020gpg}.
If $A$ and $\Gamma$ were independent of $z$, one could use the Green's function method to solve Eq.\ (\ref{eq:boltzmann1}); however, $A$ is a function of $m_i(z)/T$.  To deal with this dependence on $z$, we  discretize space, $z\rightarrow z_0 + n\Delta z$ with
$n=0,\cdots, N-1$, and  Fourier transform Eq.\ (\ref{eq:boltzmann1}),
\be
\frac{2\pi i}{\Delta z} \left(\frac{k}{N}-\left\lfloor\frac{2k}{N}\right\rfloor\right) \Tilde{q}_k + \frac{1}{N}\sum\limits_{l=0}^{N-1}\widetilde{\left(A^{-1}\Gamma\right)}_{(k-l)\mathrm{\,mod\,} N} \,\Tilde{q}_l= \widetilde{(A^{-1}S)}_k,\quad k=0,\cdots, N-1,
\label{FTboltz}
\ee
where the tilde denotes the discrete Fourier transform. This is a linear system that is straightforward to numerically solve for $\Tilde{q}_k$. Once $\Tilde{q}_k$ is known, it can be
transformed back and interpolated to obtain $q(z)$. Eq.\  (\ref{eq:boltzmann2}) can then be integrated using a Runge-Kutta algorithm.\\

Finally, one can substitute Eq.\ (\ref{eq:perturbation}) into the Higgs EOM (\ref{eq:EOMs}) to express the friction in terms of the fluid perturbations $\mu_i$, $\delta\tau_i$ and $u_i$. This leads to the result
\be
\int \frac{d^3p}{(2\pi)^3 2E}\,\delta f_i = \frac{T_+^2}{2}\left[ C_0^{1,0}\mu_i + C_0^{0,0}(\delta\tau_i+\delta\tau_\mathrm{bg})+D_v^{0,-1}(u_i+u_\mathrm{bg})\right],
\ee
where the functions $C_v^{m,n}$ and $D_v^{m,n}$ can be found in Ref.\ \cite{Laurent:2020gpg}.

\section{Cosmological signatures}
\label{sec:sig}
We have now established the machinery needed to compute all the
relevant properties of the first order phase transition bubbles, starting from the fundamental parameters of the microscopic Lagrangian.
In this section we describe how to apply these results for the estimation of GW spectra and the baryon asymmetry.

%
\subsection{Gravitational Waves}
\label{sec:GW}
%

We follow the methodology of Refs.\ 
\cite{Caprini:2019egz,Hindmarsh:2017gnf,espinosa2010energy,Guo:2020grp,Hindmarsh:2020hop} to estimate future gravitational wave detectors' sensitivity to the GW signals that can be produced by a first-order electroweak phase transition in the models under consideration.
The GW spectrum $\Omega_{\mathrm{gw}}(f)$ is the contribution per frequency octave to the energy density in gravitational waves, {\it i.e.,} $\int \Omega_{\rm gw}\, d\ln f$ is the fraction of energy density compared to the critical density of the universe. The
spectrum gets separate contributions from the scalar fields, sound waves in the plasma and magnetohydrodynamical turbulence created by the phase transition:
\begin{equation} \label{spectrum}
\Omega_{\mathrm{gw}}(f)=\Omega_\phi(f)+\Omega_{\mathrm{sw}}(f)+\Omega_{\mathrm{m}}(f)\,,
\end{equation}
Each of these contributions depends on the wall velocity $v_w$, the supercooling parameter $\alpha$ (Eq.\ (\ref{alphadef})), and the inverse duration of the phase transition, defined as
\begin{equation}
\beta=H(T_n)T_n\left.\frac{d}{dT}\frac{S_3}{T}\right|_{T=T_n}.
\end{equation}
Another useful quantity is the mean bubble separation, which can be written in terms of $v_w$ and 
$\beta$ as~\cite{Caprini:2019egz}
\begin{equation}
R=\frac{(8\pi)^{1/3}}{\beta}\max[c_s,v_w].
\label{eq:typicalRadius}
\end{equation}
It has been shown in Ref.\ \cite{Bodeker:2017cim} that interactions with gauge bosons prevent the wall from running away indefinitely towards $\gamma\rightarrow\infty$. In that case, the contribution from the scalar fields has been shown to be negligible. Furthermore, the estimates for the magnetohydrodynamical turbulence are very uncertain and sensitive to the details of the phase transition dynamics~\cite{Pol:2019yex}, and are expected to be much smaller than the contribution from sound waves. Hence, we consider only the effects from the latter, and set $\Omega_\mathrm{m}(f)=\Omega_\phi(f)=0$.  For convenience, we reproduce the numerical fits of the GW spectra derived in Refs.\ \cite{Caprini:2019egz,Hindmarsh:2017gnf,espinosa2010energy,Guo:2020grp,Hindmarsh:2020hop} in appendix \ref{app:gw}.\\
     
We will use these predictions with respect to four proposed space-based GW detectors: LISA \cite{Audley:2017drz}, AEDGE \cite{Bertoldi:2019tck}, BBO \cite{Corbin:2005ny} and DECIGO \cite{Seto:2001qf}. A successful GW detection depends upon having a large enough signal-to-noise ratio \cite{thrane2013sensitivity},
\begin{equation}
\mathrm{SNR}=\sqrt{\mathcal{T} \int_{f_{\min }}^{f_{\max }} d f\left[\frac{\Omega_{\mathrm{gw}}(f)}{ \Omega_{\mathrm{sens}}(f)}\right]^{2}}
\end{equation}
where $\Omega_{\mathrm{sens}}(f)$ denotes the sensitivity of the detector\footnote{For AEDGE, we use the envelope of minimal strain that can be achieved by each resonance,
with its width scaled to approximate $\Omega_\mathrm{sens}(f)$. This curve is expected to reproduce the correct SNR up to about 10\%.} and $\mathcal{T}$ is the duration
of the mission. The sensitivity curves for the detector LISA, BBO and DECIGO were obtained from Ref.\ \cite{Breitbach:2018ddu}. Whenever $\mathrm{SNR}$ is greater than a given threshold $\mathrm{SNR}_{\mathrm{thr}}$, we conclude that the signal can be detected. In general, this threshold can depend upon the configuration of the detector. For all the experiments, we take $\mathrm{SNR}_\mathrm{thr}=10$ and $\mathcal{T}=1.26\times10^8\ \mathrm{s}$. In the following,  $\mathrm{SNR}_{\rm max}$ will designate the maximum signal-to-noise ratio detected by one of the detectors:
\be
\mathrm{SNR}_{\rm max}\equiv\max[\mathrm{SNR}_\mathrm{LISA},\,\mathrm{SNR}_\mathrm{AEDGE},\,\mathrm{SNR}_\mathrm{BBO},\,\mathrm{SNR}_\mathrm{DECIGO}].
\ee\\

While $\Omega_\mathrm{sens}(f)$ can be obtained from the noise spectrum of a detector, it is not practical to compare it to the GW spectrum directly; one needs to compute the SNR to determine if a signal is detectable. A useful tool for visualizing the sensitivity of a detector is the peak-integrated sensivity curve (PISC) defined in Refs.\ \cite{Alanne:2019bsm,Schmitz:2020syl,Schmitz:2020rag}, which is a generalization of the power-law sensitivity curve~\cite{Thrane:2013oya}. The main advantage of the former is that it does not assume a power-law spectrum, hence it conserves all the information about the SNR. In the simple case where one considers the contribution from only one GW source, the PISC can be obtained by factorizing the GW spectrum as
\be
\Omega_\mathrm{gw}(f)=\Omega_\mathrm{p}\, S(f,f_\mathrm{p}),
\ee
where $f_\mathrm{p}$ and $\Omega_\mathrm{p}=\max[\Omega_\mathrm{gw}(f)]$ are the peak frequency and GW amplitude and $S$ is a function that parametrizes the spectrum's shape, with a maximum at $f=f_\mathrm{p}$ and $S(f_\mathrm{p},f_\mathrm{p})=1$. One can then write the SNR as 
\be
\mathrm{SNR} = \mathrm{SNR}_\mathrm{thr}\,\frac{\Omega_\mathrm{p}}{\Omega_\mathrm{PISC}(f_\mathrm{p})},
\ee
with the PISC
\be
\Omega_\mathrm{PISC}(f_\mathrm{p})=\mathrm{SNR}_\mathrm{thr}\left[\mathcal{T}\int_{f_{\min }}^{f_{\max }} df\left(\frac{S(f,f_\mathrm{p})}{\Omega_\mathrm{sens}(f)}\right)^2\right]^{-1/2}.
\ee
By construction, any GW signal that peaks above the PISC has $\mathrm{SNR} > \mathrm{SNR}_\mathrm{thr}$ and can therefore be detected. 

%
\subsection{Baryogenesis}
\label{sec:EWBG}
%

The mechanism of electroweak baryogenesis is sensitive to the speed and shape of the bubble wall during the phase transition.
In most previous studies, these quantities were treated as
free parameters to be varied, but in this work we have  already derived them, as was discussed in Section \ref{sect:shape}. 
An important requirement for EWBG is to
avoid the washout, by baryon-violating sphaleron interactions, of the generated asymmetry inside the bubbles of broken phase, once they have formed.  This leads to the well-known constraint
\cite{moore1998measuring}
\be
{v_n\over T_n}>1.1\,,
\label{sphwash}
\ee
which was derived within the SM for low Higgs masses where a first order EWPT was possible.  The bound can be slightly higher (up to 1.2) in singlet-extended models \cite{fuyuto2014improved}, depending upon the parameters, due to the sphaleron energy
being modified.  Here we adopt 
the SM constraint (\ref{sphwash}); we checked that taking the
more stringent bound 1.2 removes $\sim 5\%$ of viable models in the scan over parameter space to be described below.\\

Near the bubble wall, CP-violating processes associated with the effective interaction
in Eq.\ (\ref{top_quark_mass}) give rise to
perturbations of the plasma, that result in a local chemical potential $\mu_{B_L}$ for left-handed baryons, which by imposing the chemical equilibrium of strong-sphaleron interactions, is related to those of the $t_L$, $t_R^c$ and $b_L$ quarks by 
\begin{equation}
\mu_{B_{L}}=\frac{1}{2}\left(1+4 K_{1}^{t}\right) \mu_{t}+\frac{1}{2}\left(1+4 K_{1}^{b}\right) \mu_{b}-2 K_{1}^{t} \mu_{t^{c}}\,,
\label{mubleq}
\end{equation}
where the $K_1^{a}$ functions were defined in~\cite{Fromme:2006wx} ($K_1^a = D^a_0$ in the notation of  \cite{Cline:2020jre}).
The $\mu_{B_L}$ potential biases sphalerons, leading to baryon number violation, whose associated Boltzmann equation can be integrated to obtain the 
baryon to photon ratio\footnote{The extra factor of $\gamma_w = 1/\sqrt{1-v_w^2}$ in the denominator was pointed out by Ref.\ \cite{Cline:2020jre}.}
\begin{equation}
\eta_{b}=\frac{405\, \Gamma_{\mathrm{sph}}}{4 \pi^{2} v_{w}
\gamma_w g_{*} T} \int d z\, \mu_{B_{L}} f_{\mathrm{sph}} e^{-45 \Gamma_{\mathrm{sph}}|z| / 4 v_{w}}\,,
\end{equation}
where $f_{\rm sph}$ quantifies the
diminution of the sphaleron rate in the broken phase \cite{DOnofrio:2014rug,cline2011electroweak}.
The most challenging step for the computation of EWBG is in the determination of the chemical potentials $\mu_{t_L}$, $\mu_{t^c_R}$ and $\mu_{b_L}$ appearing in Eq.\ (\ref{mubleq}).
They satisfy fluid equations resembling the network
(\ref{eq:boltzmann1},\ref{eq:boltzmann2}), except that the potentials relevant for EWBG are CP-odd, whereas those determining the wall profiles are CP-even.  \\

The CP-odd transport equations have been discussed extensively in the literature, leading to two schools of thought as to how best to compute the source term for the CP asymmetries.  These are commonly known as the VEV-insertion \cite{Riotto:1995hh,Riotto:1997vy} or WKB (semiclassical) \cite{Joyce:1994zt,Cline:2000nw,Kainulainen:2001cn,Kainulainen:2002th,Prokopec:2003pj,Prokopec:2004ic} methods, respectively.  A detailed discussion and comparison of the two approaches was recently given in Ref.\ \cite{Cline:2020jre}, which quantified the well-known fact that the VEV-insertion source tends to predict a larger  baryon asymmetry than the WKB source, by a factor of $\sim 10$.  In the present work we adopt the WKB approach, which was updated in Ref.\ \cite{Cline:2020jre} to allow for consistently treating walls moving near or above the sound speed.  In addition, that reference computed the source term arising from the same effective interaction (\ref{top_quark_mass}) as in the present model, so we can directly adopt the CP-odd fluid equations studied there.

%
\section{Monte Carlo results}
\label{sec:MC}
%

To study the properties of the phase transition, we performed a scan over the parameter space of the models, imposing several constraints. We found that variations in $\lambda_s$ do not qualitatively change the results, prompting us to initially fix its value at $\lambda_s=1$, leaving $\lambda_{hs}$ and $m_s$ as the free scalar potential parameters. We will first discuss this slice of parameter space, and later consider the quantitative 
dependence on $\lambda_s$.
We also chose $\Lambda=540\ \mathrm{GeV}$, which is conservative since there are no collider constraints on its value for singlet masses in the region $m_s=[110,160]$ GeV. Recall that $\Lambda$ is important for the determination of the baryon asymmetry $\eta_b$, which is expected to scale roughly as $1/\Lambda$. Finally, in order to prevent Higgs invisible decays, we imposed $m_s>m_h/2$.\\

We used a Markov Chain Monte Carlo algorithm to  efficiently explore the regions of parameter space having desired phase transition properties. Starting with an initial model
satisfying the sphaleron bound (\ref{sphwash}), one
generates a new trial model by randomly varying the parameters
$\lambda_i$ by small increments $\delta_i$. The trial model is added to the chain using a conditional 
probability
\be
   P=\min\left[\frac{v_n/T_n}{1.1},1\right]
   \label{probdef}
\ee
that favors models having strong first order phase transitions,
and for which a solution to the nucleation condition
 (\ref{nucleation})  can be found.
 We adjust the $\delta_i$ so that roughly half of the models are kept in successive trials, with larger values of $\delta_i$ being more likely to result in a rejection.\\  
 
\begin{figure}[t]
	\centering
	\includegraphics[width=0.31\textwidth]{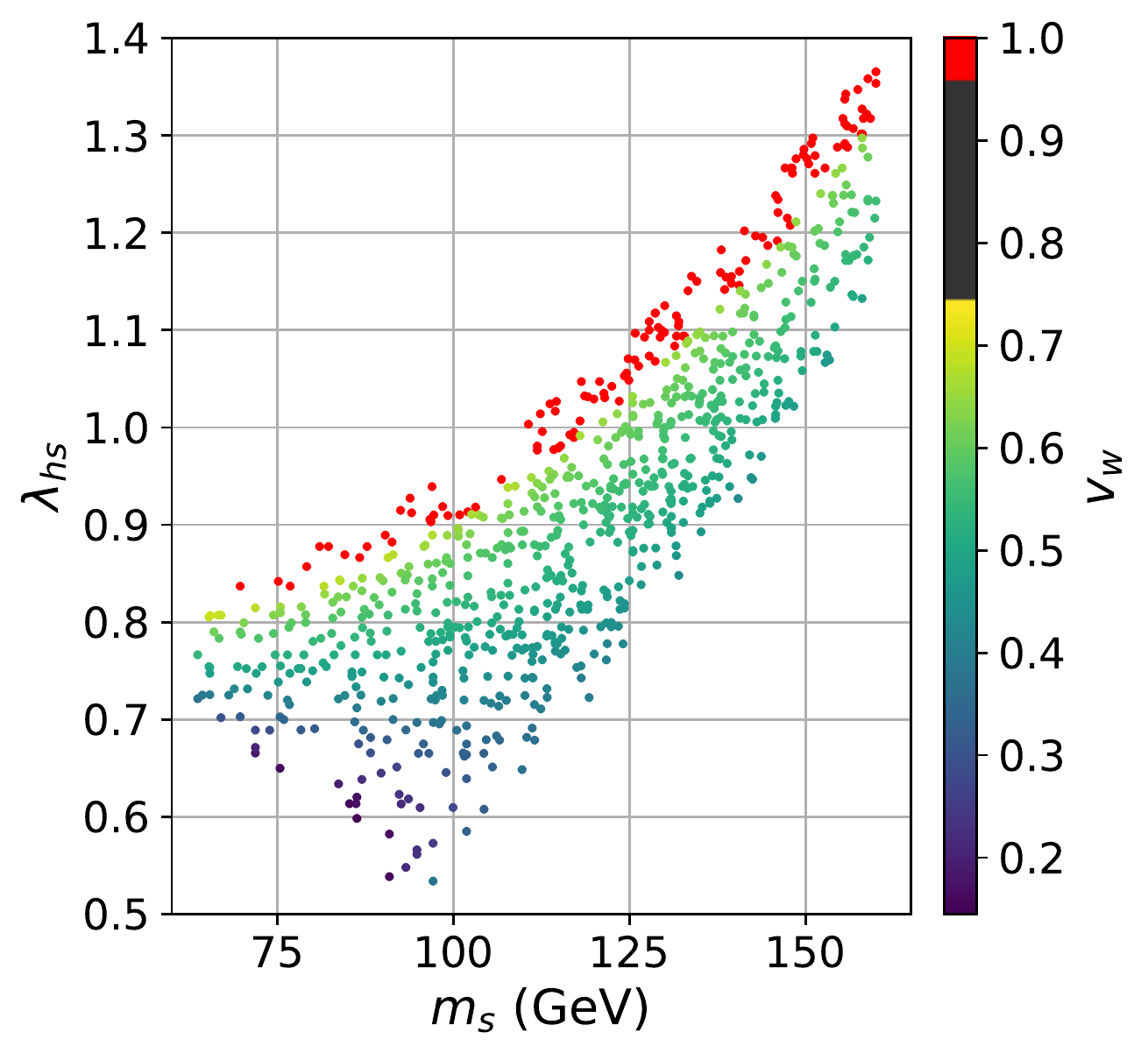}\hspace{2mm}\includegraphics[width=0.32\textwidth]{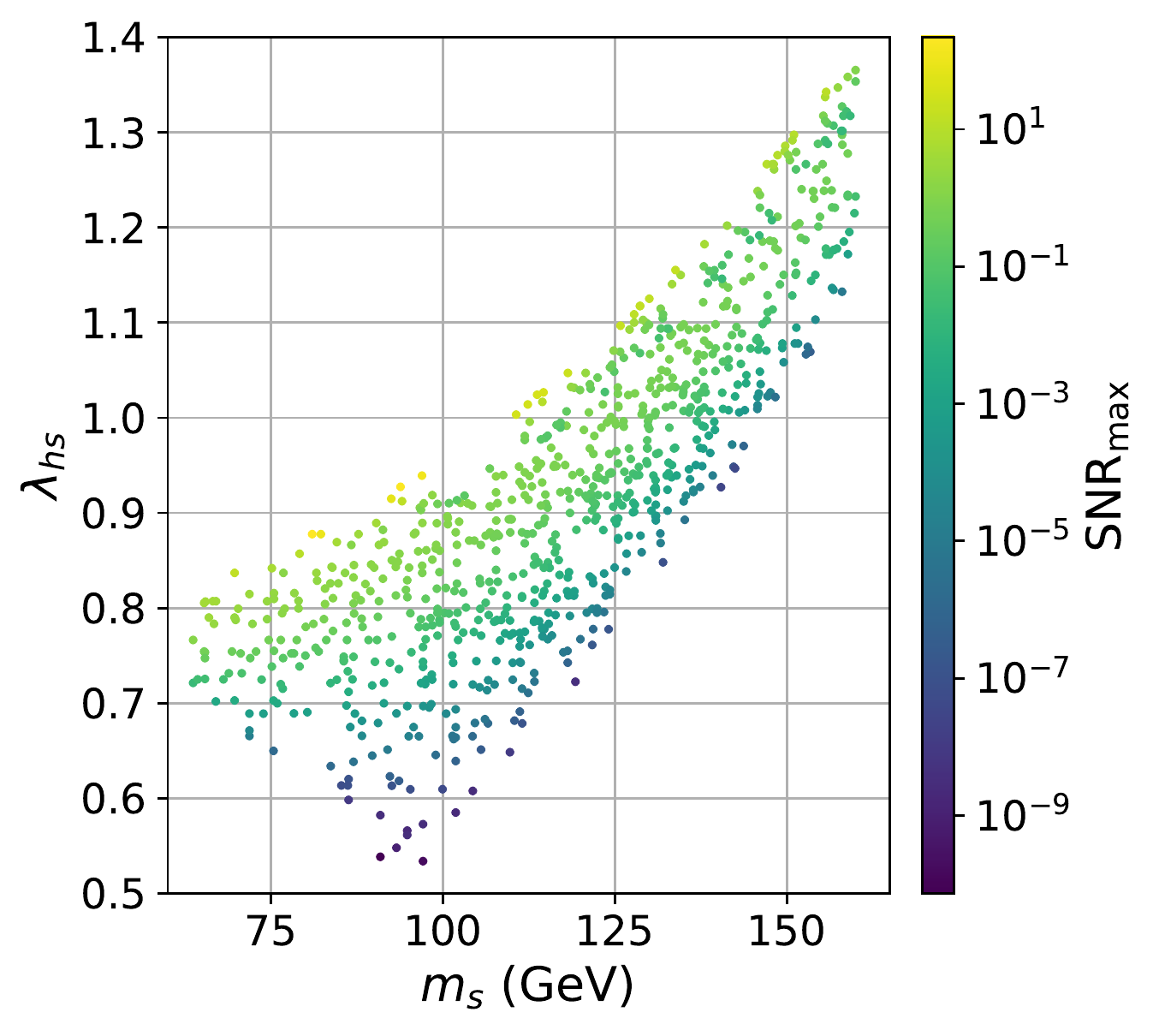}\hspace{2mm}\includegraphics[width=0.32\textwidth]{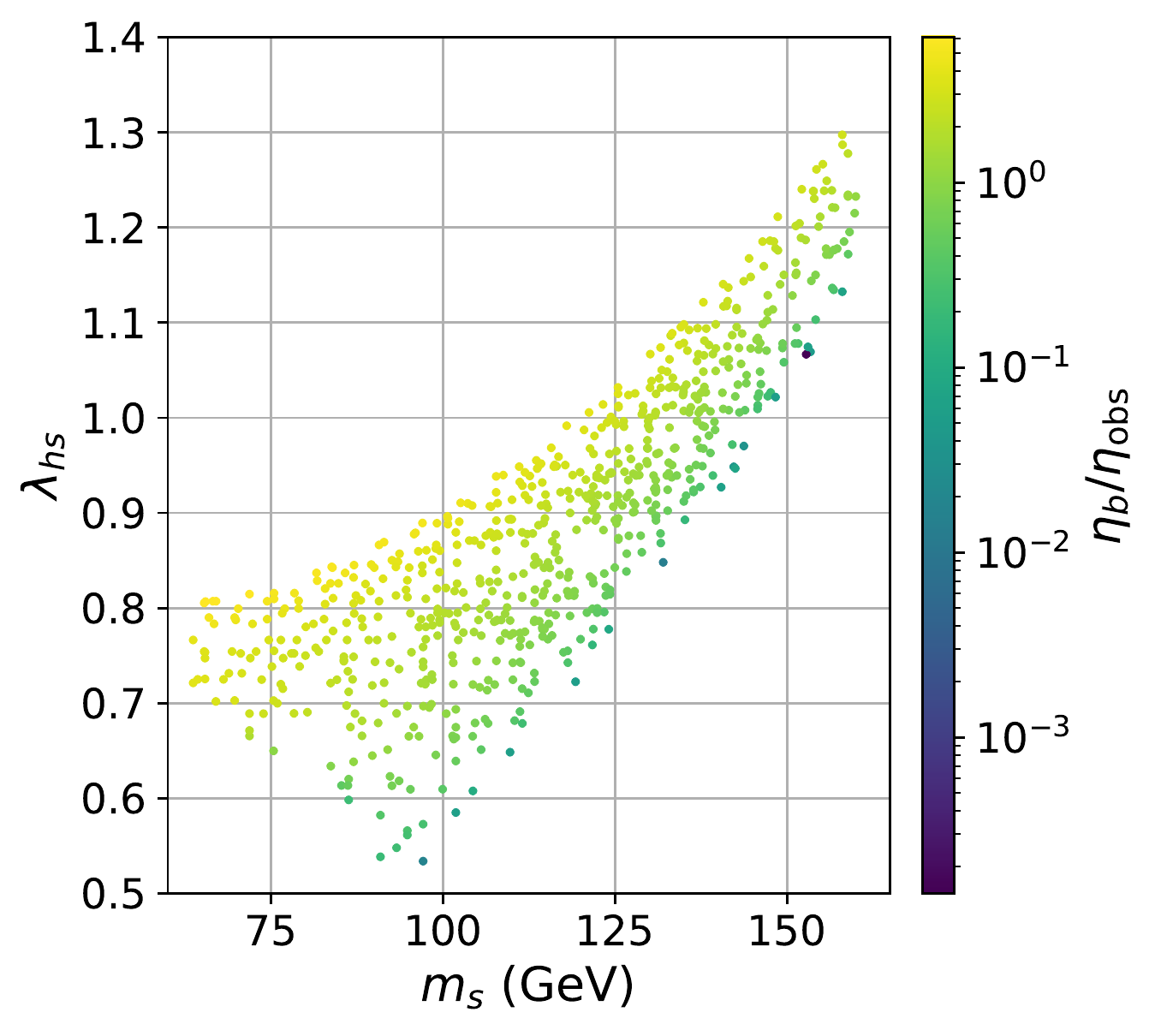}
	\centerline{ (a) \qquad\qquad\qquad\qquad\qquad\qquad\qquad\qquad (b) \qquad\qquad\qquad\qquad\qquad\qquad\qquad\qquad (c)}
	\caption{\small{Scan of the parameter space with $\lambda_s=1$ and $\Lambda=540\ \mathrm{GeV}$. The colors represent (a) the terminal wall velocity $v_w$, (b) the maximum signal-to-noise ratio of gravitational waves that could be detected by either LISA, AEDGE, BBO or DECIGO and (c) the baryon asymmetry (in units of the observed value) produced by the phase transition. The red dots in (a) correspond to detonation solutions with $v_w\approx1$, and the
	latter are not included in (c) since they are expected to produce a negligible baryon asymmetry (see text).}}
	\label{fig:paramspace}
\end{figure}

 This procedure yielded 842 models with strong phase transitions,
 of which 712 were amenable to finding solutions for the moment
 equations (\ref{eq:moment1}-\ref{eq:moment2}). Our analysis typically works for $\gamma\lesssim10$; for faster walls, the algorithm for determining the wall properties becomes numerically unstable and does not yield reliable results. This is due to the large
 ($500\times 500)$ matrix $\widetilde{\left(A^{-1}\Gamma\right)}$ of eq.\ (\ref{FTboltz}) becoming singular as $v_w\to 1$.  It is therefore difficult to determine the type of solution of the 130 remaining models using our methodology alone: they could either stabilize at ultrarelativistic speeds, or (from a naive perspective---see below) run away indefinitely towards $\gamma\rightarrow\infty$. The value of the baryon asymmetry should not be affected by this ambiguity since it is negligible for $v_w\approx1$. The GW spectrum produced during the phase transition is sensitive to this distinction since runaway walls have a nonnegligible fraction of their energy stored in the wall, while for non-runaway walls, the energy gets dissipated into the plasma, so the fraction of energy in the wall becomes negligible. This ambiguity can be lifted using the result of Ref.\ \cite{Bodeker:2017cim}, which found that in the limit $\gamma\rightarrow\infty$, interactions between gauge bosons and the wall create a pressure proportional to $\gamma$, preventing it from running away.\footnote{More recently, the authors of Ref.~\cite{Hoeche:2020rsg} have carried out an all-orders resummation at leading-log acuracy, finding that the pressure is in fact proportional to $\gamma^2$ for fast-moving walls.}\ \ We therefore assume that the 130 models without a solution to the moment equations (\ref{eq:moment1}-\ref{eq:moment2}) correspond to non-runaway walls with $v_w\approx1$. 
 The results of this scan, showing the calculated wall velocity, signal-to-noise ratio of gravity waves observable by at least one of the proposed experiments (LISA, AEDGE, BBO or DECIGO), and the predicted baryon asymmetry (in units of the observed value) are presented in Fig.\ \ref{fig:paramspace}, in the plane of of $\lambda_{hs}$ versus $m_s$.\\
 
\subsection{Deflagration versus detonation solutions}

\begin{figure}[t]
	\centering
	\includegraphics[width=0.43\textwidth]{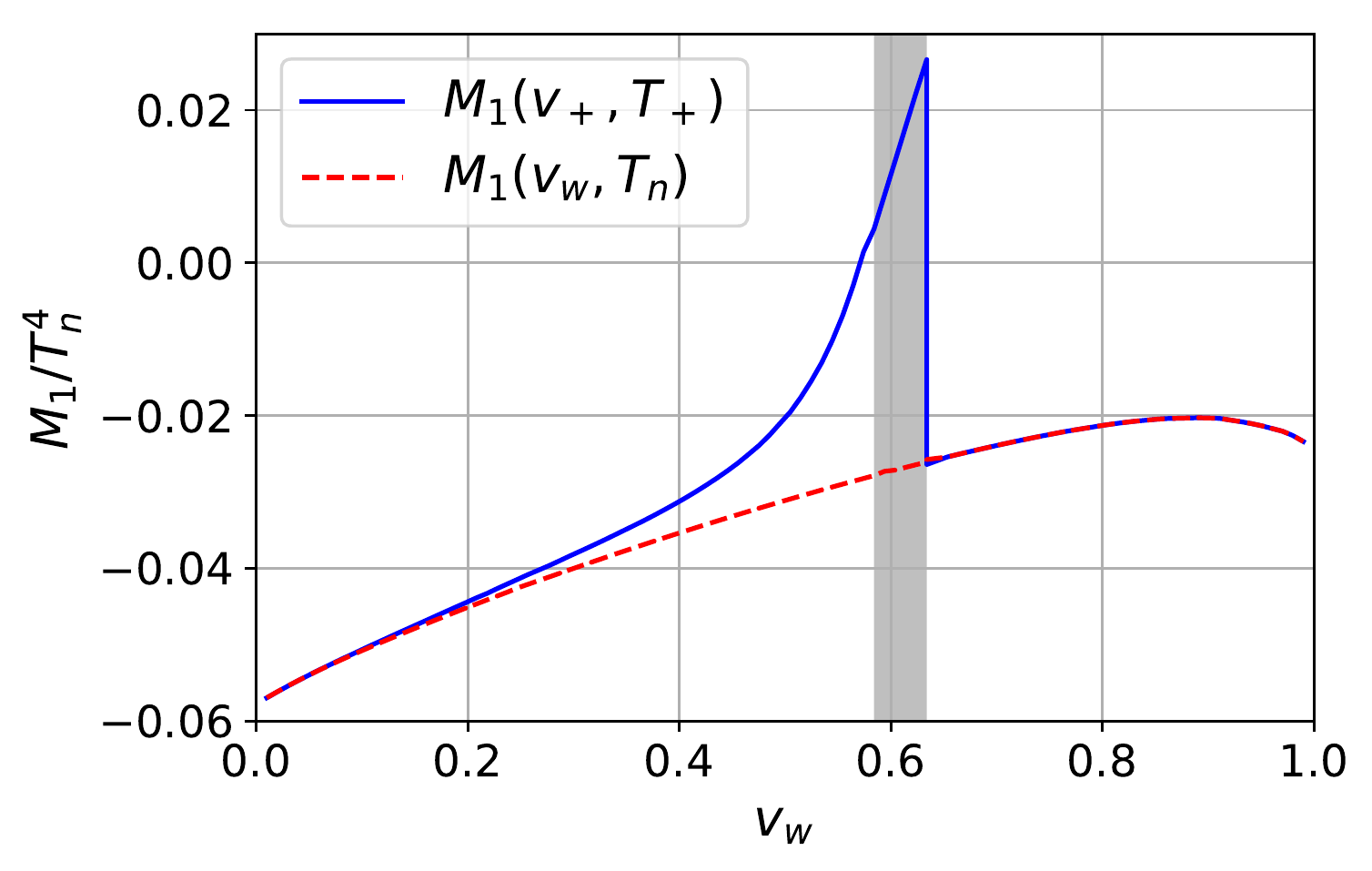}\hspace{8mm}\includegraphics[width=0.4\textwidth]{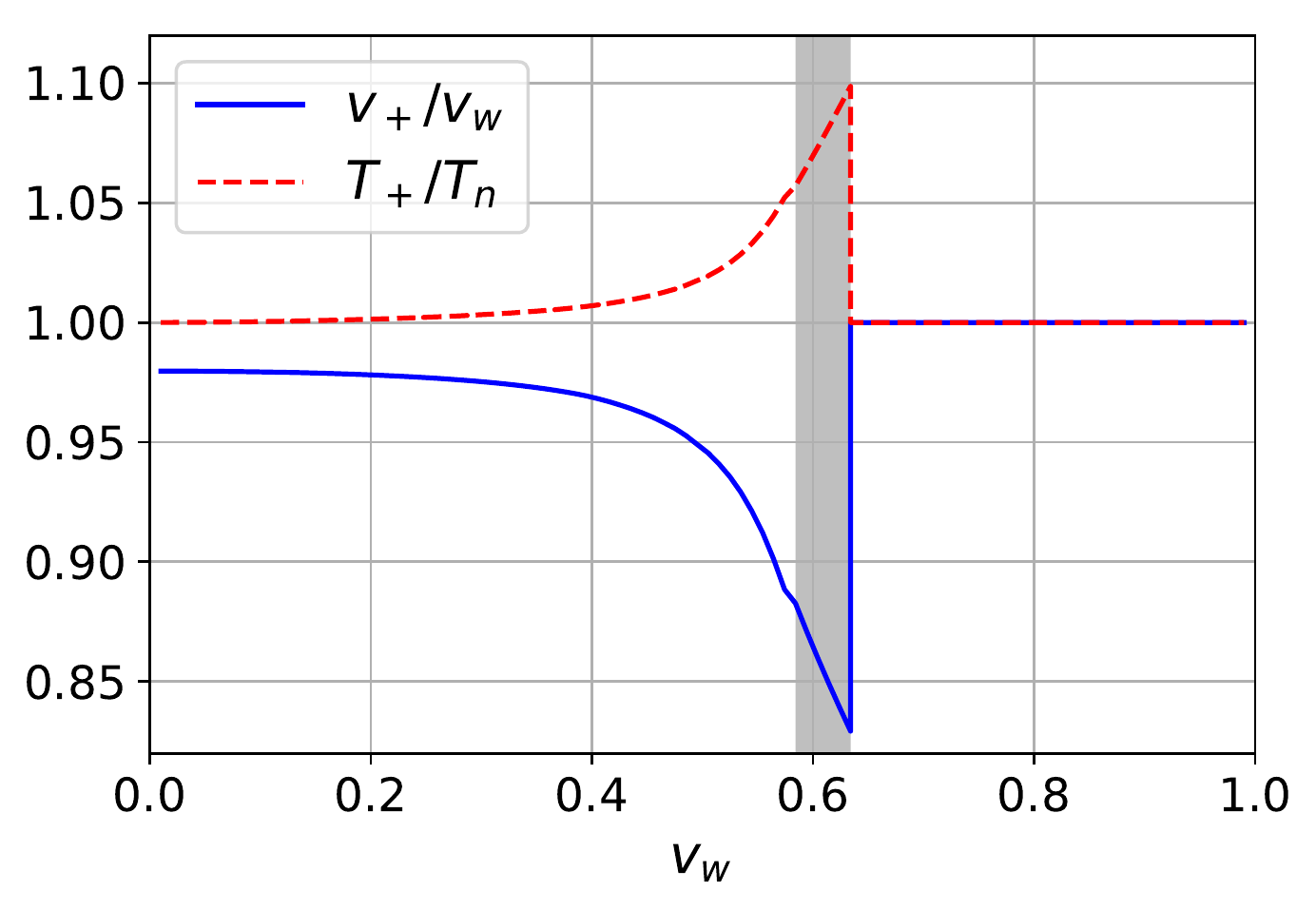}
	\centerline{\qquad (a) \qquad\qquad\qquad\qquad\qquad\qquad\qquad\qquad\qquad\qquad\qquad (b)}
	\caption{\small{Left (a): Pressure on the wall $M_1$ as a function of the wall velocity $v_w$. The solid (dashed) line corresponds to the pressure evaluated at the velocity $v_+$ ($v_w$) and the temperature $T_+$ ($T_n$). Right (b): Relation between the naive variables $v_w$, $T_n$ and the ones relevant for evaluating $M_1$, namely $v_+$ and $T_+$. Both plots were obtained using the parameters $m_s=130\ \mathrm{GeV}$, $\lambda_{hs}=\lambda_s=1$ and $L_h=5/T_n$. The shaded region corresponds to hybrid wall solutions characterized by $c_s<v_w<\xi_J$.}}
	\label{fig:M1}
\end{figure}

A striking feature of these results is that all the detonation solutions
have $v_w\approx1$.\footnote{Strictly speaking there are models with $v_w<1$ detonation solutions but these always have another solution at a lower velocity corresponding to a deflagration or hybrid wall. Then only the latter solution is physically relevant,  since the bubble is created at $v_w=0$ and accelerates until it reaches the solution with the lowest velocity.}\ \  We have tested that this is not
specific to the choice of fixed parameter values, but also holds for all models having $0.01<\lambda_s<8$ and $\Lambda>110\ \mathrm{GeV}$; hence it seems to be a general property of phase transitions in the $Z_2$-symmetric singlet framework. One can understand this behavior by considering the net pressure opposing the wall's expansion, $M_1$ (recall Eq.\ (\ref{eq:moment1}-\ref{eq:moment2})), as a function of the wall velocity, as illustrated in Fig.\ \ref{fig:M1}.  It shows how $M_1$ differs when evaluated with the appropriate quantities $v_+, T_+$ rather than the incorrect ones $v_w, T_n$.  Using the latter, we would find no solution to the equation $M_1=0$ for the exemplary model used in Fig.\ \ref{fig:M1}, and would then incorrectly conclude that it satisfies $v_w\approx1$. The relevant quantities are those measured right in front of the wall, $v_+$ and $T_+$. The speed $v_+$ is smaller than $v_w$ for $v_w<\xi_J$, which would lower the pressure against the wall ($\xi_J$ is the Jouguet velocity, defined as the smallest velocity a detonation solution can have). However, in the same region, the temperature $T_+$ is larger than $T_n$, which causes the pressure to increase. The latter effect turns out to dominate over the former. Indeed, the actual pressure, represented by the solid blue line in Fig.\ \ref{fig:M1}, increases much more rapidly than $M_1(v_w,T_n)$ close to the speed of sound. This qualitative difference allows for a solution to $M_1=0$, which would have been missed if we had used the naive quantities $v_w$ and $T_n$.\\

\begin{figure}[b]
	\centering
	\includegraphics[width=0.32\textwidth]{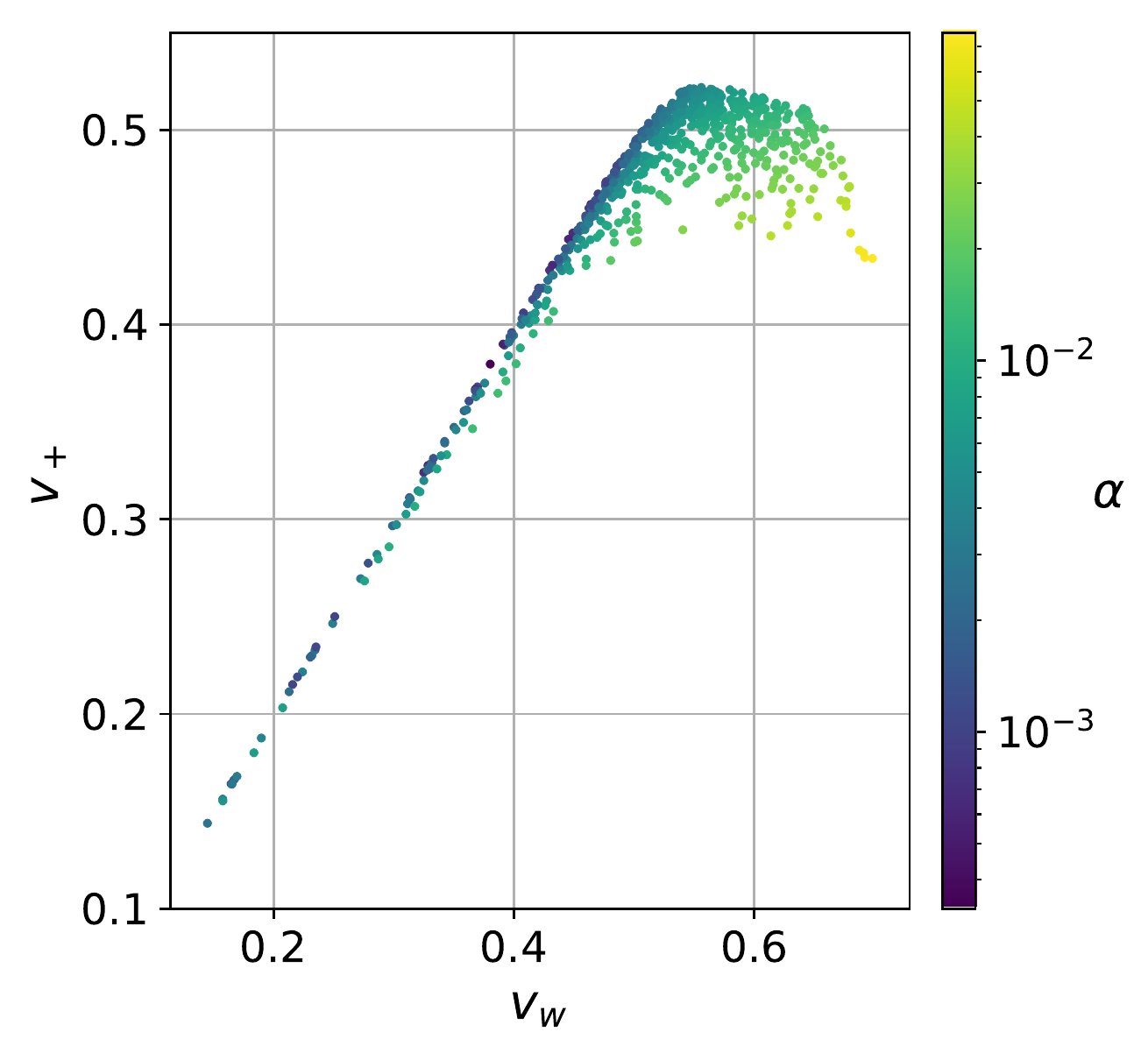}\hspace{2mm}\includegraphics[width=0.32\textwidth]{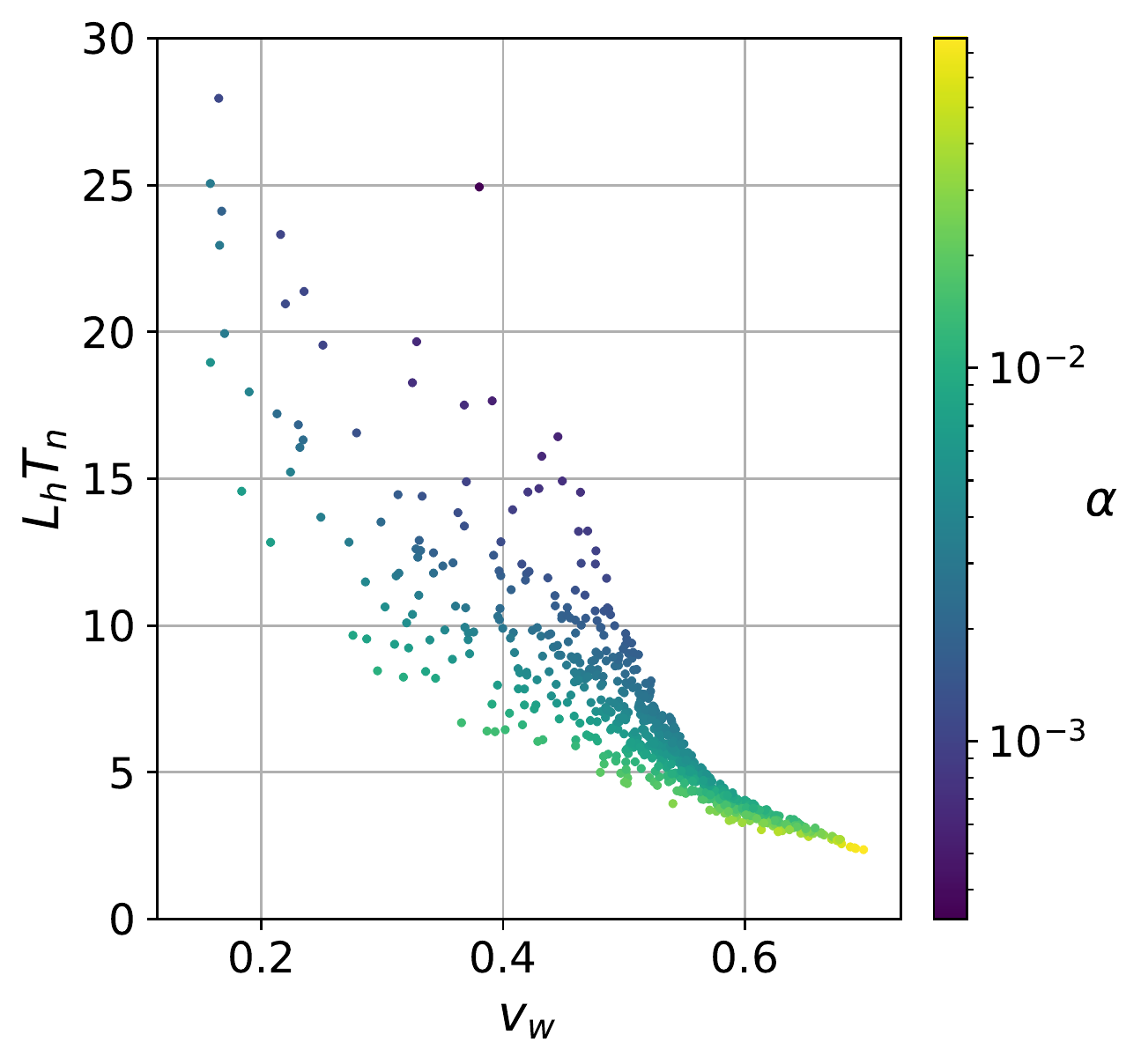}\hspace{2mm}\includegraphics[width=0.32\textwidth]{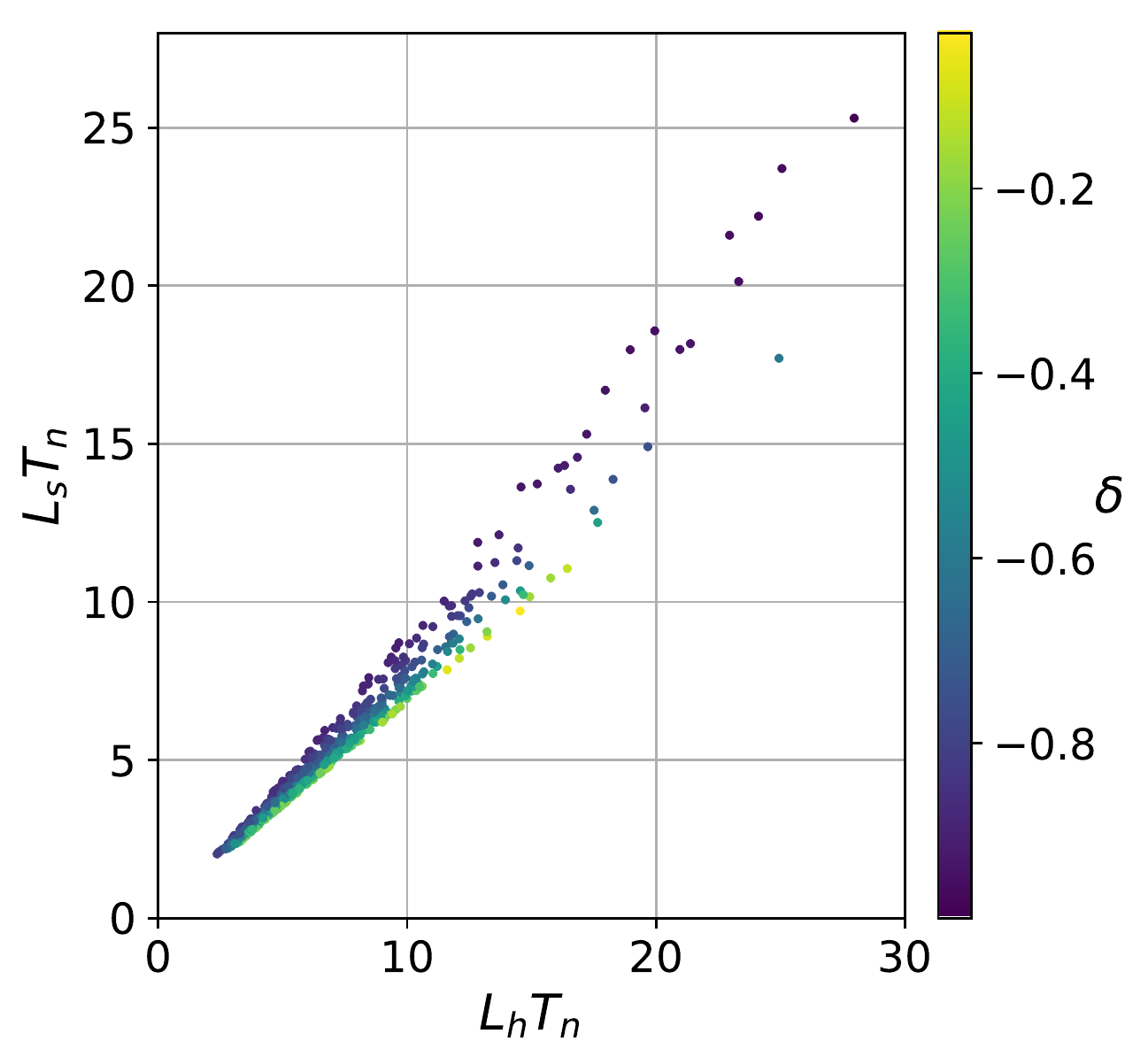}
	\centerline{ (a) \qquad\qquad\qquad\qquad\qquad\qquad\qquad\qquad (b) \qquad\qquad\qquad\qquad\qquad\qquad\qquad\qquad (c)}
	\caption{\small{Shape and velocity of the deflagration solutions.   (a) Correlation between  the wall velocity $v_w$ and the fluid velocity in front of the wall, $v_+$; (b) dimensionless wall width $L_h\times T_n$ versus $v_w$;  and (c) correlation of the $s$ and $h$ wall widths.  Colors indicate the supercooling parameter $\alpha$ (Eq.\ (\ref{alphadef})) in (a,b), or the wall offset $\delta$
	(Eq.\ (\ref{shapedef})) in (c).}}
	\label{fig:wallshape}
\end{figure}

We find that the previous statements apply quite
generally: for all models, $T_+>T_n$ when $v_w<\xi_J$, and this always leads to a much higher pressure on the wall, even if the difference between $T_+$ and $T_n$ is quite small; the pressure barrier at $v_w=\xi_J$ is always greater than the maximum possible value for a detonation solution. Therefore, if the phase transition is strong enough to overcome the pressure barrier at $\xi_J$, the solution becomes a detonation, but the pressure in the region $v_w>\xi_J$ is never enough to prevent it from accelerating towards $v_w\approx1$. If the phase transition is weaker, the pressure barrier is high enough to impede the detonation, and it becomes a deflagration or hybrid solution.\\

The wall thickness and speed for the models with deflagration\footnote{Henceforth we take ``deflagration'' to also
include hybrid solutions} solutions are shown in Fig.\ \ref{fig:wallshape}, which demonstrates that the behaviors for subsonic (deflagration) and supersonic (hybrid) walls are qualitatively different.  Subsonic walls generally have $v_+\approx v_w$, which is  expected since the fluid should not be strongly perturbed by a slowly moving wall. The wall width is not uniquely determined by $v_w$, but there exists a clear correlation, with slower walls being thicker. For supersonic cases, the correlation between $v_+$ and $v_w$ gets inverted: higher wall velocity leads to lower $v_+$. The wall width becomes uniquely determined by $v_w$ and the relation between these two variables is to a good approximation linear. One observes that stronger phase transitions, quantified by  higher values of $\alpha$, generally produce faster and thinner walls. Even for the strongest transitions our solutions still have wall thickness $LT\gsim 3$. Since the semiclassical force mostly affects particles with momenta $\langle k_z\rangle  \sim T$, we find $L \langle k_z\rangle \gsim 3$, so that the semiclassical approximation is still valid. In fact the semiclassical picture has been shown to remain valid for surprisingly narrow walls~\cite{Jukkala:2019slc}, working very well for $L \langle k_z\rangle \approx 4$ and still reasonably for $L \langle k_z\rangle \approx 2$. There is a linear correlation between the $h$ and $s$ wall widths, but the slope is not 1; in all cases, we find that $L_h>L_s$.  The distribution of wall offset values $\delta$ is also indicated in Fig.\ \ref{fig:wallshape}(c).\\

\begin{figure}[t]
	\centering
	\includegraphics[width=0.32\textwidth]{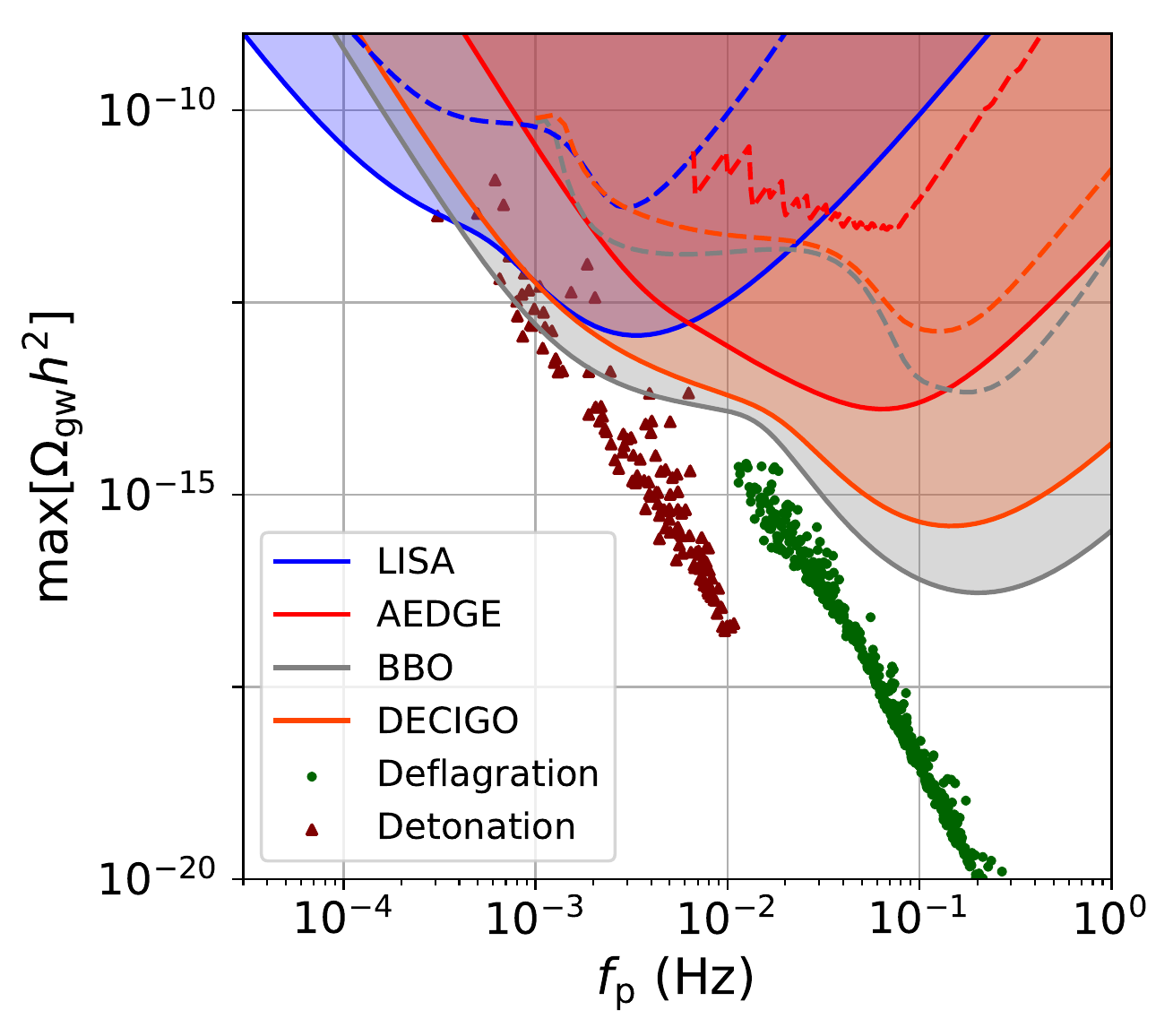}\hspace{2mm}\includegraphics[width=0.32\textwidth]{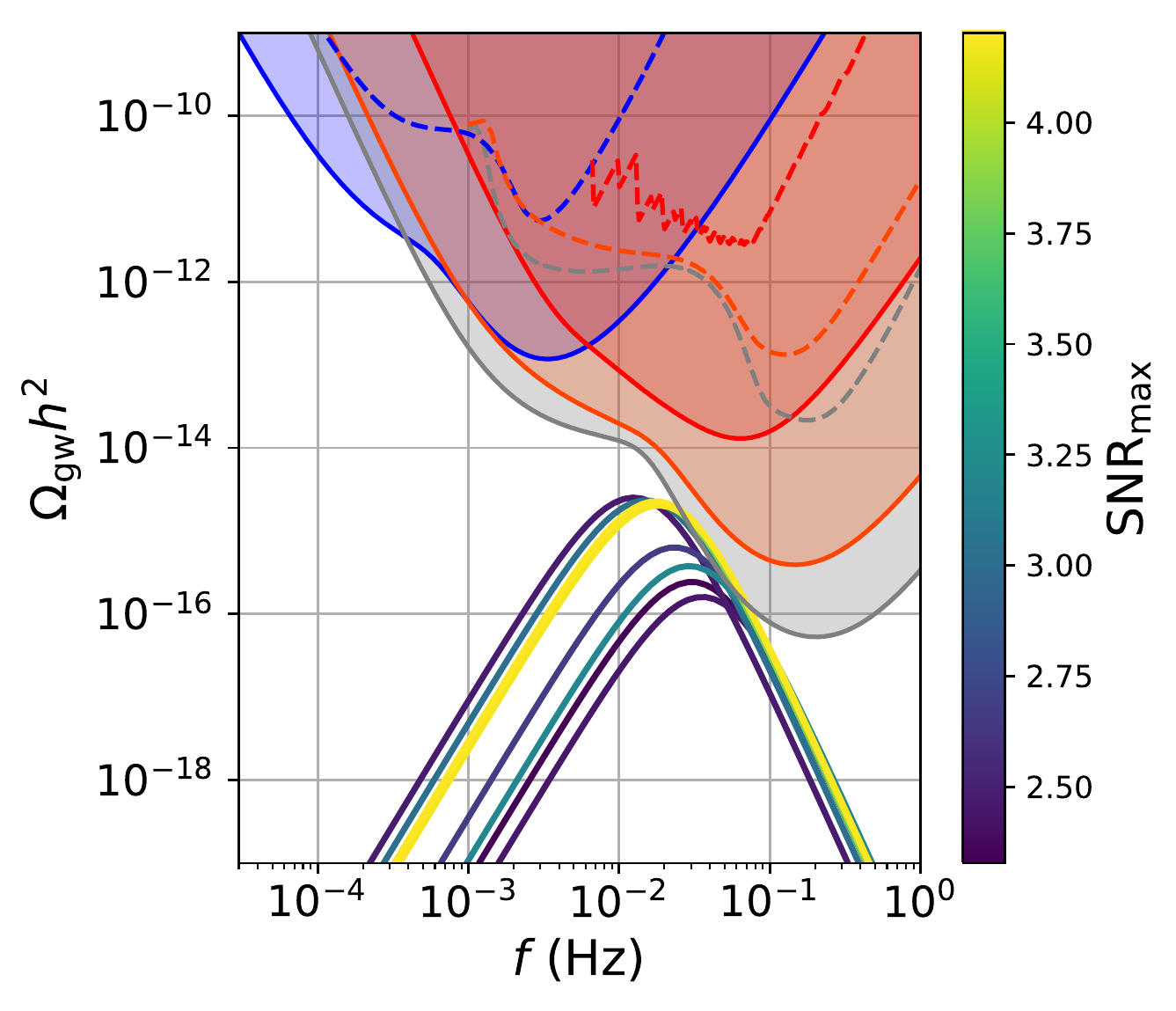}\hspace{2mm}\includegraphics[width=0.32\textwidth]{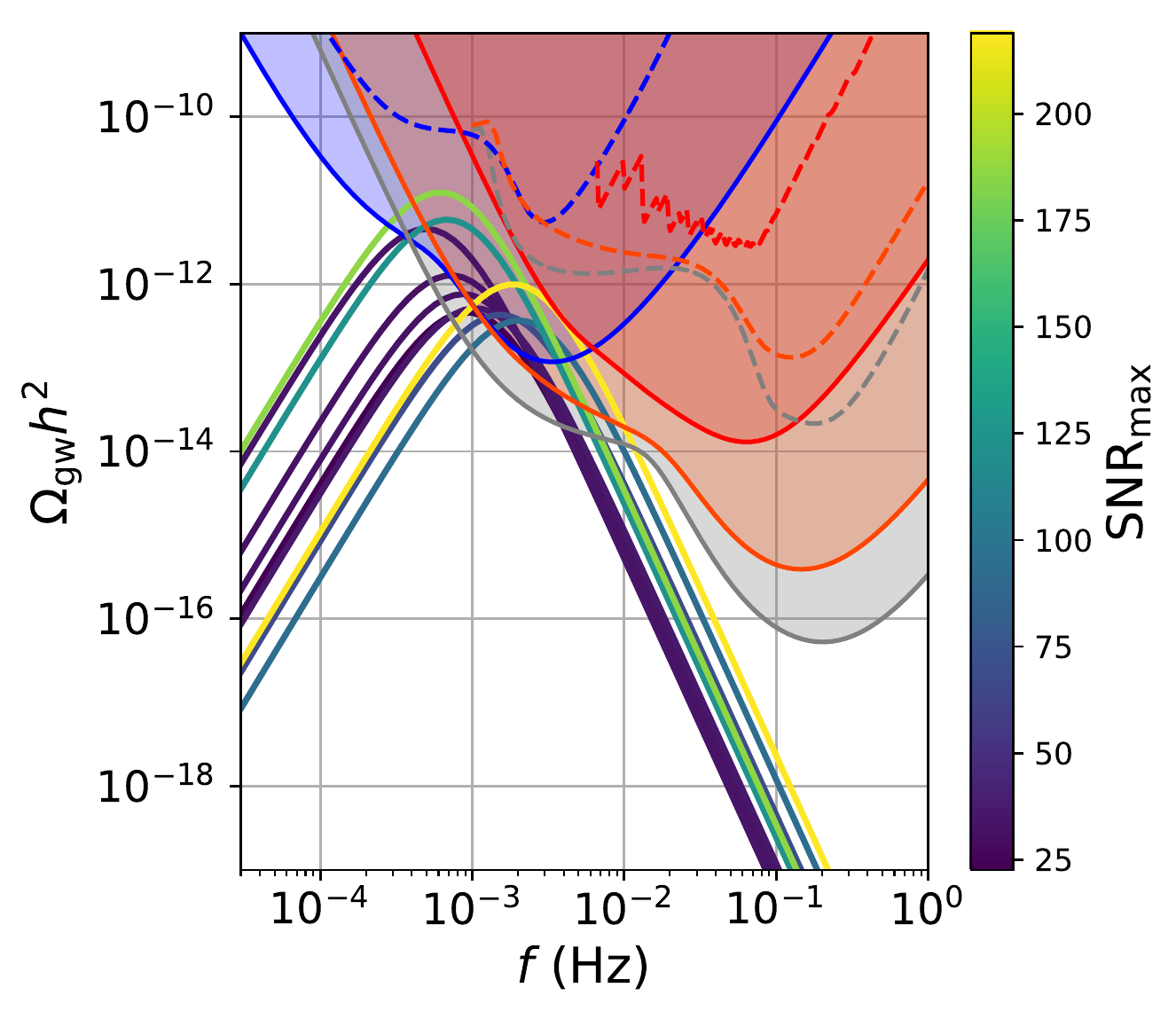}
	\centerline{ (a) \qquad\qquad\qquad\qquad\qquad\qquad\qquad\qquad (b) \qquad\qquad\qquad\qquad\qquad\qquad\qquad\qquad (c)}
	\caption{\small{(a): Maximum amplitude of GW as a function of the peak frequency $f_\mathrm{p}$ with the peak-integrated sensitivity curve $\Omega_\mathrm{PISC}h^2$ (solid line) and the sensitivity $\Omega_\mathrm{sens}h^2$ (dashed line) of the four considered detectors. (b) and (c): Spectrum of GWs produced by the 10 models with the highest $\mathrm{SNR}_\mathrm{max}$ for (b) deflagration and (c) detonation solutions.}}
	\label{fig:GWSpectrum}
\end{figure}

\subsection{Baryogenesis and gravity wave production}

Of the 842 sampled models, 517 are able to generate the baryon asymmetry at a level large enough to agree with observations, and 20 detonation walls can produce observable gravitational waves. We found no detectable deflagration solutions. More detailed results are presented in Table \ref{table:scans}. The complementarity of the experiments considered here,
with respect to the present model, can be appreciated by considering the relation between the maximum GW amplitude\footnote{$h=0.678$ is the reduced Hubble constant defined by $H_0=100h\, \mathrm{km\, s^{-1}\, Mpc^{-1}}$ \cite{Aghanim:2018eyx}.} $\max[\Omega_\mathrm{gw}h^2]$ and the frequency of this peak amplitude $f_\mathrm{max}$, as shown in Fig.\  \ref{fig:GWSpectrum}\,(a). The peak frequency of the strongest detonation walls are positioned exactly in LISA's region of maximal sensitivity, while the peak frequency of the deflgration solutions are closer to the peak sensitivity of AEDGE, DECIGO and BBO. The complete spectrum's shape are also shown in Fig.\  \ref{fig:GWSpectrum}\,(b,c)  for deflagration and detonation solutions respectively. We conclude that detonation walls could be probed by LISA, DECIGO and BBO, but not by AEDGE.\\

\begin{figure}[t]
	\centering
\centerline{	\includegraphics[width=0.45\textwidth]{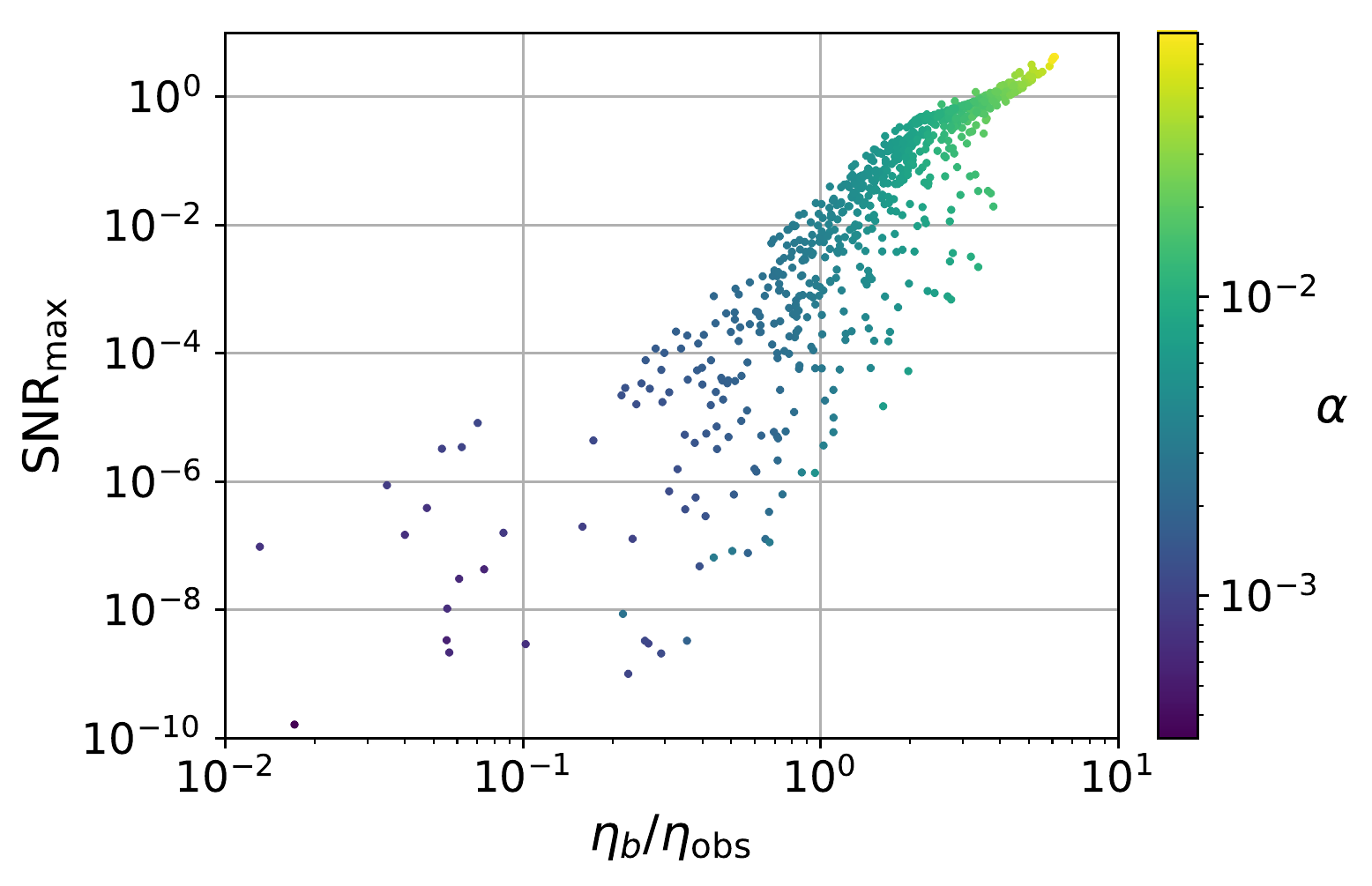}\hspace{5mm}\includegraphics[width=0.45\textwidth]{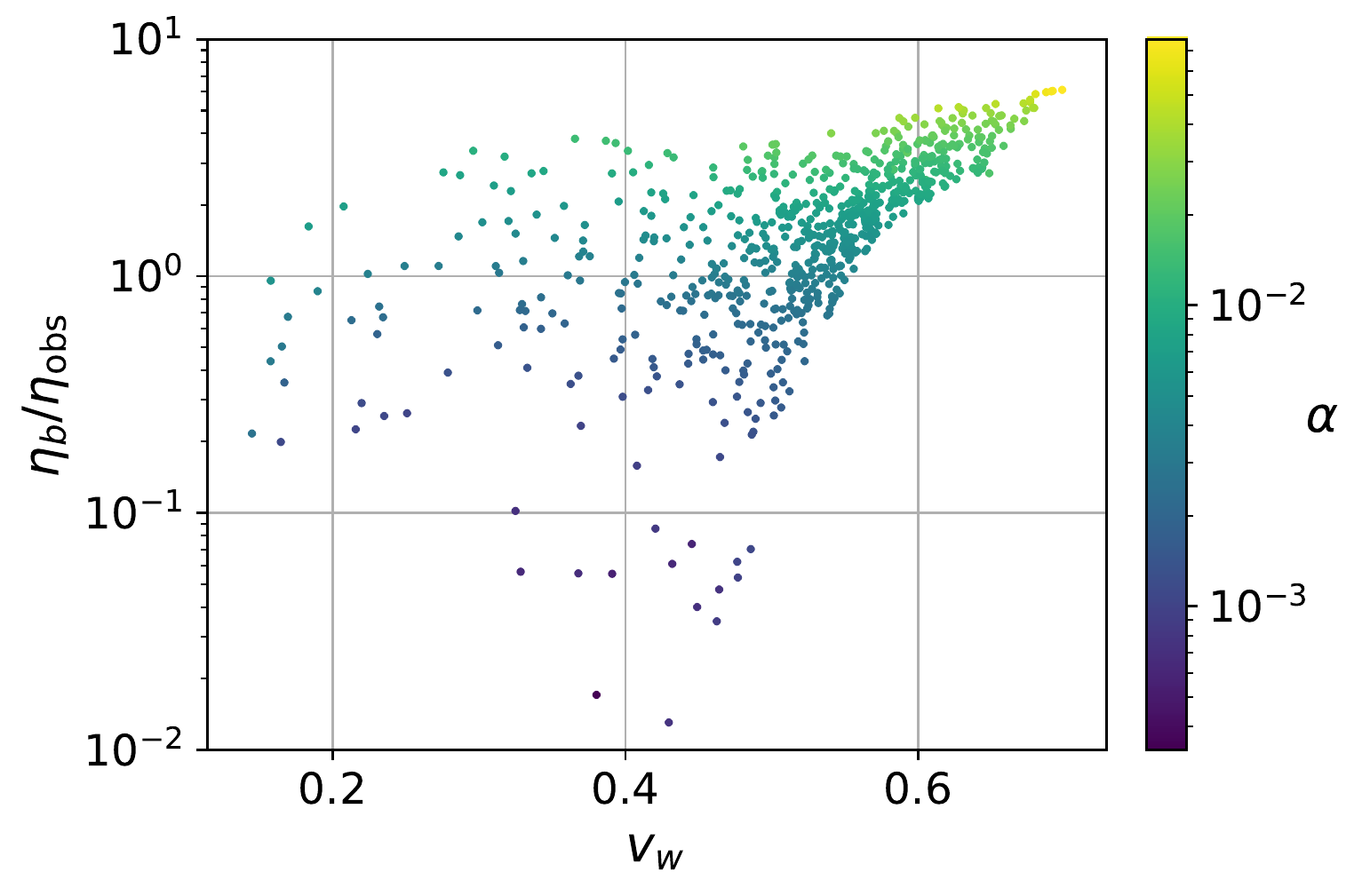}}
\vskip-0.5cm
	\leftline{ \qquad\qquad (a) \hspace{8.5cm}(b)}
	\caption{\small{(a): Relation between the $\mathrm{SNR}_\mathrm{max}$ and the baryon asymmetry produced by the phase transition. (b): Baryon asymmetry as a function of the wall velocity. Both plots only show the deflagration models.}}
	\label{fig:etaSNR}
\end{figure}

In previous studies, where the wall velocity was considered as a free parameter, there was an expectation that 
baryogenesis would be less efficient with increasing $v_w$,
whereas gravity waves would become more so.  In the present study, where $v_w$ is not adjustable but is a derived parameter, we surprisingly find that rather than EWBG
and stronger GWs being anticorrelated, instead they are
positively correlated, as is illustrated in 
Fig.\ \ref{fig:etaSNR}\,(a).  This can be understood from the fact (see Fig.\ \ref{fig:wallshape}\,(b)) that $L_h$ is a decreasing function of $v_w$, which enhances EWBG. Moreover, the relevant velocity for EWBG is $v_+$, which is a decreasing function of $v_w$ for supersonic walls, and is bounded by $v_+<c_s$; this effect also enhances EWBG for fast-moving walls. The actual relation between $\eta_b$ and $v_w$ is shown in Fig.\ \ref{fig:etaSNR}\,(b) and, at least for supersonic walls, there is a positive correlation between these two variables. Fig.\ \ref{fig:etaSNR} also indicates that the supercooling parameter $\alpha$ is positively correlated with both $\eta_b$ and $\mathrm{SNR}_\mathrm{max}$: stronger phase transitions generally lead to both higher GW and baryon production.\\

Detailed predictions for EWBG in the $Z_2$ symmetric model were previously made in Refs.\ \cite{vaskonen2017electroweak} and \cite{Xie:2020wzn}, as opposed to merely requiring the sphaleron bound
(\ref{sphwash}) to be satisfied. Comparisons with the present work are hindered by the fact that different source terms for the CP asymmetry were assumed.  In Ref.\ \cite{vaskonen2017electroweak}, the dimension-6 coupling $i(y_t/\sqrt{2})(s/\Lambda)^2\bar h t_L t_R$ was used, rather than the dimension-5 coupling in Eq.\ (\ref{top_quark_mass}).  Moreover, a value $v_w = 0.2$ was taken for the wall velocity, and an estimate $L_h = v_n/\sqrt{8 V_b}$ was made for the wall width, where $v_n$ is the Higgs VEV at the nucleation temperature, and $V_b$ is the potential barrier between the two minima.  For the same potential parameters ($\lambda_s = 0.1$) as in \cite{vaskonen2017electroweak}, we find no values of $v_w$ below 0.43, and our determination of $L_h$ is two to three times larger than the estimate in 
\cite{vaskonen2017electroweak}.  Both of these discrepancies would lead to overestimating
the efficiency of EWBG, helping to explain why Ref.\ \cite{vaskonen2017electroweak} obtains a high frequency of successful models, despite the extra suppression that should result from using a dimension-6 source term.\\

In Ref.\ \cite{Xie:2020wzn}, the dimension-5 coupling to leptons rather than the top quark was studied, and a different formalism (the VEV insertion approximation)
for computing the CP asymmetry was employed, which tends to give significantly larger estimates for the baryon asymmetry than the WKB method that we adopt 
\cite{Cline:2020jre}.   For the parameters of the benchmark models taken in that paper, we find
significantly higher wall velocities, $v_w\sim 0.6$-$0.7$ than the values $v_w\lesssim 0.1$ that were needed to match the observed baryon asymmetry there. This can be compensated by increasing the CP-violating phase $\phi=0.02$ assumed there by a factor of $\sim 10$.  We are reanalyzing this alternative source term
within the EWBG formalism used in the present paper (work in progress).


\subsection{Dependence on $\lambda_s$ and $\Lambda$}

To study the quantitative dependence on the singlet self-coupling $\lambda_s$, we performed 3 other scans similar to the one previously described, taking $\lambda_s=\mathrm{0.01,\ 0.1\ and\ 8}$ (the largest value being near the limit of perturbative unitarity) and $\Lambda=540\ \mathrm{GeV}$. The results of these scans are summarized in Table \ref{table:scans}. We find that EWBG remains efficient for $\lambda_s\gtrsim0.1$. Again, we found no deflagration walls producing detectable GW, and no models detectable by AEDGE. These results confirm that only detonation solutions, which are not good candidates for EWBG, could be probed by GW detectors. Increasing $\lambda_s$ generally leads to stronger phase transitions, resulting in more models with successful EWBG and detectable GWs.  \\

The value of $\Lambda$ (recall Eq.\ (\ref{eq:Lambda})) can in principle also have an effect on the strength of the phase transition, through the effective potential's dependence on the top quark mass. The leading thermal term added to the potential varies like $h^2s^2T^2/\Lambda^2$, which becomes negligible at high $\Lambda$, but could significantly modify the behavior of the phase transition for $\Lambda\sim T_n$, resulting in a larger baryon asymmetry and GW production. We have verified that this term is already subdominant when $\Lambda=540\ \mathrm{GeV}$. However, for $m_s>110\ \mathrm{GeV}$, the weaker constraints allow for values of $\Lambda$ as low as 300 GeV, which could have an important effect on the phase transition.\\

To test the sensitivity to lower values of $\Lambda$, we repeated the previous scans using $\Lambda=\Lambda_\mathrm{min}(m_s)$, where $\Lambda_\mathrm{min}$ is given by
\be
\Lambda_\mathrm{min}(m_s) = \left\lbrace \begin{array}{ll} 
540\ \mathrm{GeV},\quad m_s<110\ \mathrm{GeV}\\
300\ \mathrm{GeV},\quad 110\ \mathrm{GeV} < m_s < 160\ \mathrm{GeV}
\end{array}\right.
\ee
The results  are shown in Table \ref{table:scans}\footnote{The $\lambda_s=0.01$ scan
is omitted since all accepted 
models satisfy $m_s<110\ \mathrm{GeV}$, making the results identical to those of the previous scan.}. As one could anticipate from the relation $\eta_b\sim1/\Lambda$, EWBG is more efficient at
lower values of $\Lambda$. One can also see that the number of detonation walls or walls generating detectable GW does not change substantially, which indicates that the lower values of $\Lambda$ do not change the character of the phase transition.

\begin{table}[t]
\centering

\begin{tabular}{|c|c|c|c|c|c|c|c|} 
\hline
\multirow{2}{*}{$\Lambda$} & \multirow{2}{*}{$\lambda_s$} & \multirow{2}{*}{$\eta_b/\eta_\mathrm{obs} > 1$} &  \multicolumn{5}{c|}{Detonation}  \\ 
\cline{4-8}
 & & & Total & $\mathrm{SNR}_\mathrm{max} > 10$ & $\mathrm{SNR}_\mathrm{LISA} > 10$ & $\mathrm{SNR}_\mathrm{BBO} > 10$ & $\mathrm{SNR}_\mathrm{DECIGO} > 10$ \\ 
\hline
 & 0.01 & 0 & 80.5 & 2.68 & 0.8 & 2.5 & 0.27 \\ 
\cline{2-8}
540 & 0.1 & 10.1 & 53 & 0.89 & 0.2 & 0.89 & 0.2\\ 
\cline{2-8}
GeV & 1 & $61.4_{+4.6}^{-5.6}$ & $15.4_{-1.4}^{+2.4}$ & $2.38_{\,-0}^{\,+0}$ & $0.83_{\,-0}^{\,+0}$ & $2.38_{\,-0}^{\,+0}$ & $0.71_{\,-0}^{\,+0}$\\ 
\cline{2-8}
 & 8 & 73.3 & 26.4 & 6.2 & 2.81 & 6.2 & 3.16 \\
\hline
\multirow{3}{*}{$\Lambda_\mathrm{min}$} & 0.1 & 21.6 & 49.3 & 1.39 & 0.69 & 1.19 & 0.4 \\
\cline{2-8}
 & 1 & 69.6 & 18.1 & 2.21 & 0.97 & 2.07 & 0.97 \\
\cline{2-8}
 & 8 & 85.7 & 13.8 & 3.55 & 1.01 & 3.55 & 1.52 \\
\hline
\end{tabular}
\caption{\small{Statistics from the scans performed with $\lambda_s=0.01,0.1,1,8$ and $\Lambda=540\ \mathrm{GeV\ and\ }\Lambda_\mathrm{min}$. Each entry corresponds to the percentage of models satisfying the indicated constraint. In the row for $\lambda_s=1$ and $\Lambda=540\ \mathrm{GeV}$, the exponents (indices) correspond to the error obtained by substituting the collision matrix $\Gamma$ for $2\Gamma$ ($\Gamma/2$). $\Lambda_\mathrm{min}$ is the minimum value of $\Lambda$ allowed by laboratory constraints.}}
\label{table:scans}
\end{table}

\subsection{Theoretical uncertainties}
\label{sec:theoretical_uncertaineties}

\begin{figure}[t]
	\centering
	\includegraphics[width=0.32\textwidth]{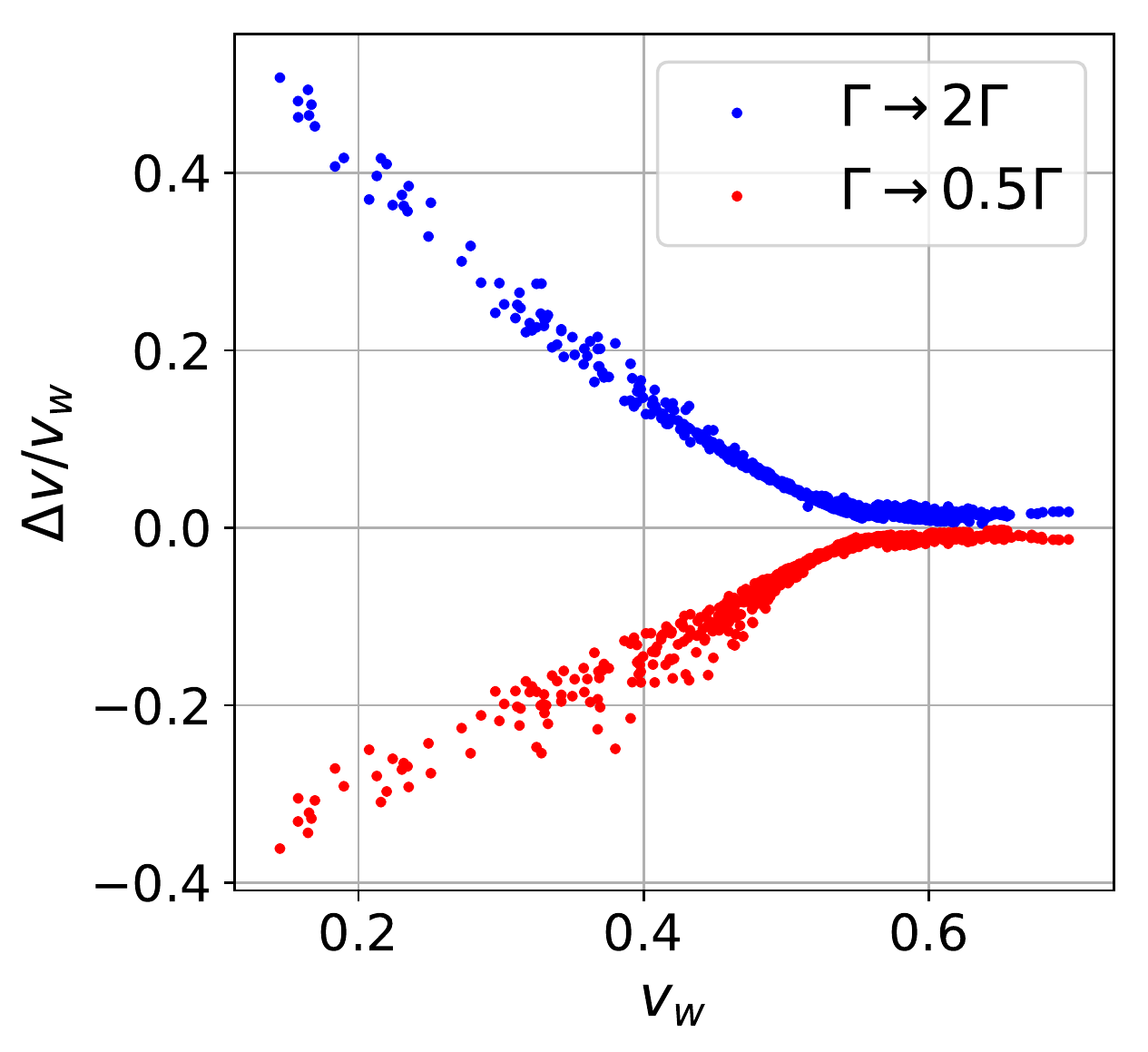}\hspace{2mm}\includegraphics[width=0.34\textwidth]{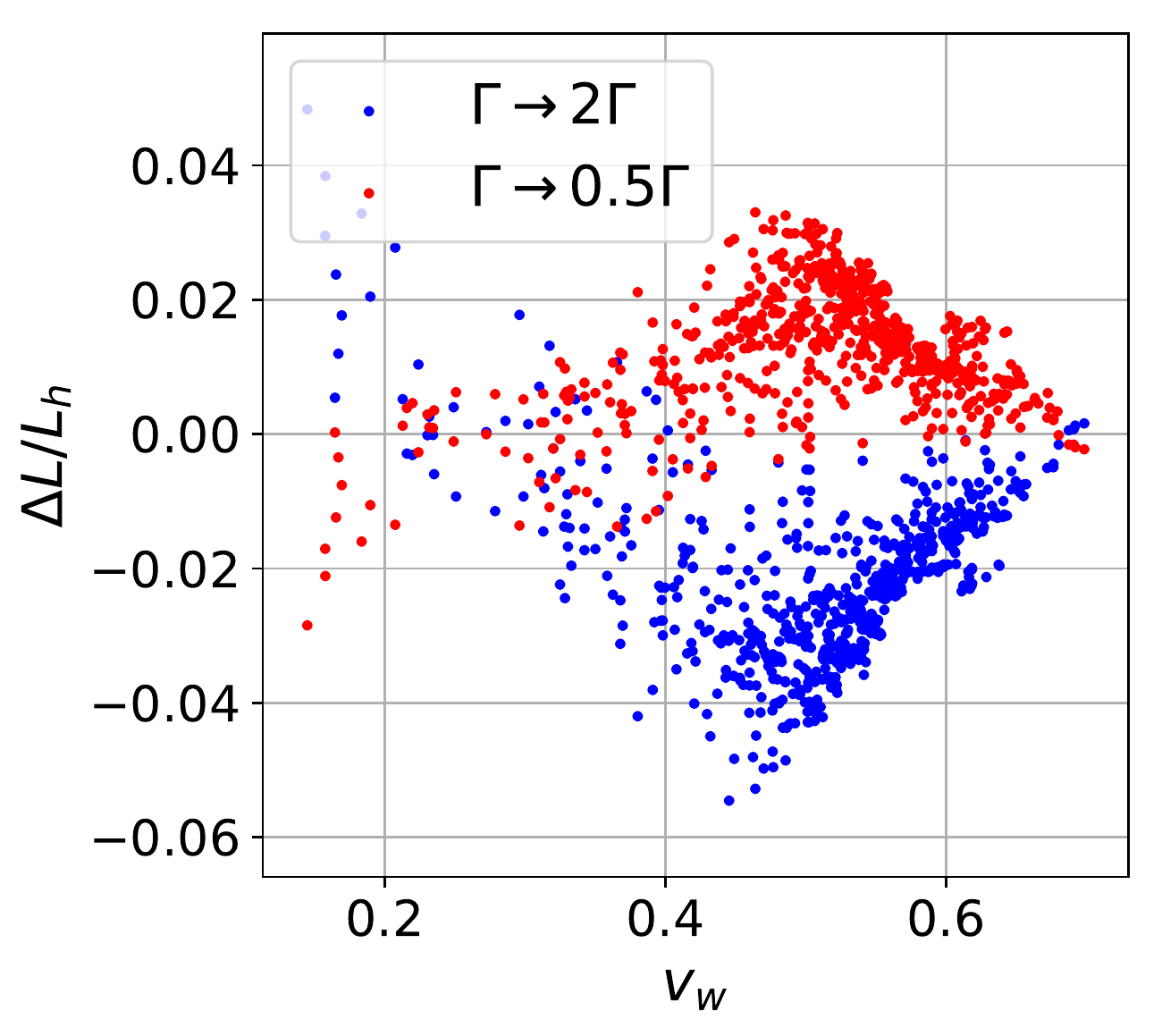}\hspace{2mm}\includegraphics[width=0.3\textwidth]{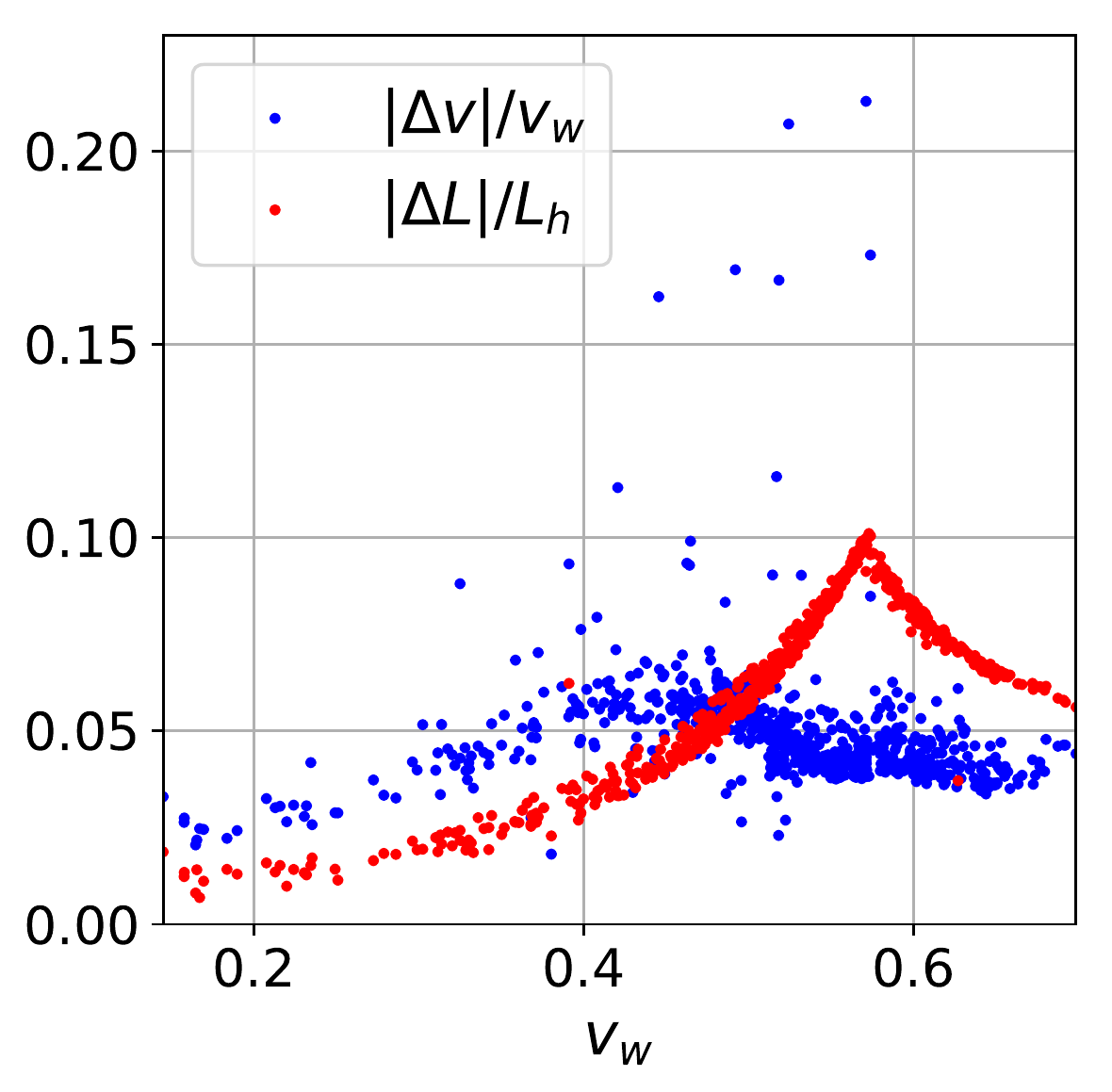}
	\centerline{ \qquad(a) \qquad\qquad\qquad\qquad\qquad\qquad\qquad\qquad\quad (b) \qquad\qquad\qquad\qquad\qquad\qquad\qquad\quad (c)}
	\caption{\small{(a) and (b): Relative changes $\Delta v/v_w$ and $\Delta L/L_h$ in the wall velocities and widths obtained by substituting $\Gamma\to 2\Gamma$ or $\Gamma/2$ respectively. (c): Absolute error on $v_w$ and $L_h$ due to the discrepancy between  the  temperatures  computed  with  the  Boltzmann equation and the conservation of the energy-momentum tensor (see Eq.\ (\ref{eq:abserrors})).}}
	\label{fig:error}
\end{figure}

In Ref.\ \cite{Laurent:2020gpg}, the  integrals that determine the collision rates $\Gamma$ appearing in the Boltzmann equation network
(\ref{eq:boltzmann1}-\ref{eq:boltzmann2}) were reevaluated, and it was noticed that the leading log approximation
that was used in their derivation leads to theoretical uncertainties of $\mathcal{O}(1)$ in the fractional error.  To study the impact of
these uncertainties on our results, 
 we recomputed the wall velocity with uniformly rescaled
 collision rates, $\Gamma\to2\Gamma$ and $\Gamma\to \Gamma/2$.
 The ensuing variations of velocity $\Delta v$ and wall width $\Delta L$ are shown in Figs.\ \ref{fig:error} (a) and (b) respectively. The effect on $v_w$ can be significant for slow walls, leading to a $\pm 40\,\%$ change when $v_w\sim 0.2$. On the other hand for nearly supersonic walls, $v_w\gtrsim c_s$, the
 wall speed is quite insensitive to $\Gamma$. The variation of $L_h$ is generally below 5\%, much smaller than the corresponding variation in $\Gamma$.\\
 
 This behavior is not surprising since, 
 near the speed of sound, the pressure on the wall is mainly determined by the variation of $T_+$, which does not depend on $\Gamma$. Likewise, the results for the baryon asymmetry and GW production turn out to be relatively robust against variations in $\Gamma$.  This is demonstrated by the error intervals in the $\lambda_s=1$ row
 of Table \ref{table:scans}. The error on the ratio of models satisfying $\eta_b/\eta_\mathrm{obs}>1$ or $\mathrm{SNR}_i>10$ is of order 10\%, which is much smaller than the range of variation in
 $\Gamma$. \\
 
 Another source of uncertainty is the discrepancy between the temperatures computed with the Boltzmann equation (see Section \ref{sec:BE}) and the conservation of the energy-momentum tensor (see Appendix \ref{app:fluidEq}). Ideally one should obtain $T_+=T_\mathrm{BE}(z\rightarrow -\infty)$ and $T_-=T_\mathrm{BE}(z\rightarrow \infty)$, where $T_\mathrm{BE}(z)=T_+(1+\delta\tau_\mathrm{bg}(z))$ is the local temperature calculated with the Boltzmann equation. The first condition is always satisfied since we impose the boundary condition $\delta\tau_\mathrm{bg}(-\infty)=0$, but we fail to recover the second one due to the different approximations made in the two methods. The discrepancy becomes larger as $v_w$ approaches the Jouguet velocity $\xi_J$, where $T_+$ increases compared to $T_-\approx T_n$ (see Fig.\ \ref{fig:M1} (b)). On the other hand, $\delta\tau_\mathrm{bg}$ does not change significantly in the same region. Hence, we observe an error in the temperature of order $\Delta T = T_- - T_\mathrm{BE}(\infty) \approx T_--T_+$.  Since the temperature is not accurate in the broken phase, the Higgs EOM is not automatically satisfied asymptotically. To solve that problem, we shift the actual Higgs VEV $h_-$ evaluated in the broken phase by an amount $-\Delta h$, so that the adjusted VEV $h_0=h_- - \Delta h$ asymptotically solves the EOM (see Eq.\ (\ref{eq:adjusted_vev})). This gives an additional source of uncertainty for $v_w$ and $L_h$.\\
 
 We estimate the errors induced on $v_w$ and $L_h$ by $\Delta T$ and $\Delta h$, assuming they are
 small enough to justify keeping just the first order terms. Assuming that $v_w$ is completely determined by the solution of $M_1=0$ and $L_h$ by $M_2=0$, the error on these solutions can be obtained by expanding around the estimated values. For example, for the error in the wall velocity is estimated by
 \be
 0=M_1(v_w+\Delta v,h_0+\Delta h,T(z)+\Delta T(z)) \approx M_1(v_w,h_0,T(z)) + \frac{\partial M_1}{\partial v_w}\Delta v + \int dz \frac{\delta M_1}{\delta T(z)}\Delta T(z) + \Delta_h M_1,
 \ee
 where $\Delta_h M_1=M_1(v_w,h_0+\Delta h,T)-M_1(v_w,h_0,T)$, and we integrate over the temperature variation because $M_1$ is a functional of $T(z)$. Since $v_w$ is the solution of $M_1(v_w,h_0,T(z))=0$, the absolute errors on $v_w$ and $L_h$ are estimated as
 \bea \label{eq:errors}
 |\Delta v| &\approx (|\Delta_T M_1| + |\Delta_h M_1|)\left\vert\frac{\partial M_1}{\partial v_w}\right\vert^{-1},\\
 |\Delta L| &\approx (|\Delta_T M_2|+|\Delta_h M_2|)\left\vert\frac{\partial M_2}{\partial L}\right\vert^{-1}\,,\\
 \eea
 where $\Delta_T M_i=\int dz ({\delta M_i}/{\delta T(z)})\Delta T(z)$. Notice that Eq.\ (\ref{eq:errors}) overestimates the errors since $\Delta_T M_i$ and $\Delta_h M_i$ have opposite signs. From Eqs.\ (\ref{eq:EOMs},\ref{eq:moment1},\ref{eq:moment2}), one can see that the functional derivative $\delta M_i/\delta T(z)$ can be approximated by $\frac{d}{dT}(\partial V_\mathrm{eff}/\partial h)$, so that
 \be
 \Delta_T M_i\approx \int dz \frac{d}{dT}\left(\frac{\partial V_\mathrm{eff}}{\partial h}\right) F_i(z)\Delta T(z)\,,
 \ee
 where $F_1=h'$ and $F_2=h'(2h-h_0)$. We can simplify this integral with the approximation $\Delta T(z)\approx (T_--T_+)[1+\tanh(z/L_h)]/2$. Furthermore, we approximate $\frac{d}{dT}\left(\frac{\partial V}{\partial h}\right)$ as being constant and half of its maximal value, occurring near $z=0$. Then
 \be
 \Delta_T M_i \approx \frac{1}{2}(T_--T_+)C_i\left.\frac{d}{dT}\left(\frac{\partial V_\mathrm{eff}}{\partial h}\right)\right\vert_{z=0},
 \ee
 where $C_1=\int dz F_1(z)[1+\tanh(z/L_h)]/2=h_0/2$ and $C_2=h_0^2/6$. Substituting this expression in Eq.\ (\ref{eq:errors}), we finally obtain that the errors on $v_w$ and $L_h$ are given by 
 \bea\label{eq:abserrors}
 |\Delta v| &\approx \left\lbrace\left\vert\frac{1}{4}(T_--T_+)h_0\frac{d}{dT}\left(\frac{\partial V_\mathrm{eff}}{\partial h}\right)\right\vert_{z=0} + |\Delta_h M_1|\right\rbrace\left\vert\frac{\partial M_1}{\partial v_w}\right\vert^{-1},\\
 |\Delta L| &\approx \left\lbrace\left\vert\frac{1}{12}(T_--T_+)h_0^2\frac{d}{dT}\left(\frac{\partial V_\mathrm{eff}}{\partial h}\right)\right\vert_{z=0} + |\Delta_h M_2|\right\rbrace\left\vert\frac{\partial M_2}{\partial L_h}\right\vert^{-1}.
 \eea\\
 
 The relative errors are presented in Fig.\ \ref{fig:error} (c) for the scan with $\lambda_s=1$ and $\Lambda=540$ GeV. The error on $v_w$ is  below 7\% for 97\% of the models, and exhibits no strong correlation with $v_w$. This happens because $\Delta T=T_--T_+$ and $dM_1/dv_w$ are roughly proportional (see Fig.\ \ref{fig:M1}), and therefore cancel each others' contributions. The relative error on $L_h$ is small at low velocity (or large $L_h$), but becomes  more significant near the speed of sound, however without ever exceeding 10\%. 
 
\subsection{Comparison of the GW signal with previous studies}
We end this section with a brief comparison with recent studies of the GW produced during a first-order electroweak phase transition. With the prospect of the upcoming LISA experiment, numerous
forecasts of the GW spectrum have been made for various extensions of the Standard Model \cite{beniwal2019gravitational,Ellis:2020nnr,Prokopec:2018tnq,Marzo:2018nov,Dev:2019njv}. Most of these find regions of model parameter space that would produce detectable GWs. Here we focus on studies of the singlet scalar extensions \cite{vaskonen2017electroweak,Kang:2017mkl,Beniwal:2017eik,Alves:2018jsw,Carena:2019une,Xie:2020wzn}.\\

Our results agree qualitatively with the conclusions of previous work, in the prediction of GWs detectable by LISA, DECIGO and BBO. However there are distinctions stemming from differences in 
methodology.
To compute the GW contribution from the sound waves, previous authors used the numerical fit presented in Ref.\ \cite{caprini2016science}, while we used the updated formulas of Refs.\ \cite{Guo:2020grp,Hindmarsh:2020hop}. This leads to a smaller peak frequency, decreasing the number of detectable models. Ref.\ \cite{caprini2016science} also does not include the factor $1-(1+2HR/\sqrt{K_\mathrm{sw}})^{-1/2}$ in the GW amplitude (see Appendix \ref{app:gw}). We find that this factor is generally quite small (of order $10^{-3}$-$10^{-2}$ for deflagrations and $10^{-2}$-$10^{-1}$ for detonations); hence the predicted GW signals are considerably reduced. \\

Another significant difference arises from our determination of the wall velocity, which was treated as a free parameter in previous work,
whereas we have computed it from the microphysics.
The GW spectrum and hence signal-to-noise ratio and ultimately the detectability are strongly dependent on the wall speed. For example, Ref.\ \cite{Kang:2017mkl} assumed $v_w=0.95$ for all models, which considerably enhanced GW production and led to more optimistic
predictions. Moreover, using a fixed value for $v_w$ hides the discontinuous transition between the deflagration and detonation solutions shown in Fig.\ \ref{fig:GWSpectrum}.

\section{Conclusion}
\label{sec:conc}

In this work we have taken a first step toward making complete predictions for baryogenesis and gravity waves from a first order electroweak phase transition, starting from a renormalizable Lagrangian that gives rise to the effective operator needed for CP-violation.  This is in contrast to previous studies in which quantities like the bubble wall velocity or thickness were treated as free parameters, instead of being derived from the microphysical input parameters as we have done here.  This is a necessary step for properly assessing the chances of having successful EWBG and potentially observable
GWs, since the two observables are correlated in a nontrivial way, when they are both computed from first principles.  \\

We have incorporated improved fluid equations, both for the CP-even perturbations that determine the friction acting on the bubble wall \cite{Laurent:2020gpg}, and for the CP-odd ones that are necessary for baryogenesis \cite{Cline:2020jre}, that can properly account for wall speeds close to the sound barrier.  Earlier versions of these equations were singular at the sound speed, making reliable predictions impossible for fast-moving walls.  Contrary to previous lore, we find that EWBG can be more efficient for faster walls, due in part to the tendency for fast walls to be thinner.\\

The $Z_2$-symmetric singlet model with vector-like top partners, analyzed in this work, was chosen for its simplicity, but the methods we used can be applied to other particle physics models that could enhance the 
EWPT.  For example, singlets with no $Z_2$ symmetry
have additional parameters, and would thus be likely to have more freedom to simultaneously yield large GW production and sufficient baryogenesis.  It would be interesting to identify other UV-completed models with these properties.
A limitation we identified with the $Z_2$-symmetric  model is that for the large values of the $\eta_2$ coupling that are desired for EWBG, the singlet self-coupling is rapidly driven toward zero by renormalization group running, above the top partner threshold.\\

{\color{black} For future work, some improvements could be made to the analysis presented here. The  wall velocity might be more accurately determined at low $v_w$ by using collision rates for the fluid perturbation equations beyond leading-log accuracy, and by including the singlet and Higgs out-of-equilibrium (friction) contributions.
Another limitation is that the current state-of-the-art for predicting the GW spectrum is
subject to large systematic uncertainties for wall velocities close to the speed of sound. Since a large fraction of deflagration transitions have $0.5\lsim v_w\lsim\xi_J$, our analysis of the GW production could greatly benefit from more accurate fits in that range of wall speeds.}\\

{\bf Acknowledgments.}  We thank T.\ Flacke, 
H.\ Guo, M.\ Lewicki, K.\ Schmitz, G.\ Servant and  K.-P.\ Xie for helpful correspondence.  The work of JC and BL was supported by the Natural Sciences and Engineering Research Council (Canada). The work of BL was also supported by the Fonds de recherche Nature et technologies (Qu\'ebec). The work of KK was supported by the Academy of Finland grant 31831.

\begin{appendix}

\section{Effective Potential}\label{app:ep}
We describe here the full effective potential used to describe the phase transition in the $Z_2$-symmetric singlet model. It takes the general form
\be \label{eq:potential}
V_\mathrm{eff}(h,s,T)=V_\mathrm{tree}(h,s)+V_\mathrm{CW}(h,s,T)+V_T(h,s,T)+\delta V(h,s).
\ee
$V_\mathrm{tree}$ is the scalar degrees of freedom's tree-level potential obtained in the unitary gauge by setting in Eq.\ (\ref{model}) $H\rightarrow h/\sqrt{2}$ and by omitting the $V_{BG}$ term:
\be
V_\mathrm{tree}(h,s)=\frac{\mu_h^2}{2}h^2+\frac{\lambda_h}{4}h^4+\frac{\lambda_{hs}}{4}h^2s^2+\frac{\mu_s^2}{2}s^2+\frac{\lambda_s}{4}s^4.
\ee
$V_\mathrm{CW}$ is the Coleman-Weinberg potential in the $\overline{\mathrm{MS}}$ renormalization scheme that incorporates the vacuum one-loop corrections and $V_T$ is the thermal potential:
\bea
V_\mathrm{CW}(h,s,T)&=\frac{1}{64 \pi^{2}} \sum_{i=W, Z, \gamma_L, 1,2, \chi, t} n_{i} \tilde{\mathscr{M}}_{i}^{4}\left(h,s,T\right)\left[\log \frac{\tilde{\mathscr{M}}_{i}^{2}\left(h,s,T\right)}{\mu^{2}}-C_{i}\right],\\
V_T(h,s,T) &= \sum_{i=W, Z, \gamma_L, 1,2, \chi, t} \frac{n_{i} T^{4}}{2 \pi^{2}} \int_{0}^{\infty}dy\, y^{2} \log \left[1\pm e^{-\sqrt{y^{2}+\mathscr{M}_{i}^{2}(h,s,T) / T^{2}}}\right] - \frac{\Tilde{g}\pi^2 T^4}{90},
\eea
where the sums go over all the massive particles, including the thermal mass. Here, we include the contribution from the W and Z gauge bosons, the photon's longitudinal polarization $\gamma_L$, the Goldstone bosons $\chi$, the top quark and the eigenvalues of the mass matrix of the Higgs boson and singlet scalar $m_1$ and $m_2$. We impose the renormalization energy scale as $\mu=v$, where $v=246\ \mathrm{GeV}$ is the Higgs vacuum expectation value. The $\pm$ in the thermal integral is $+$ for fermion and $-$ for bosons and $\Tilde{g}=\sum\limits_B N_B+\frac{7}{8}\sum\limits_F N_F = 85.25$ with the sums running over all the lighter degrees of freedom not included in the first term of $V_T$. The $C_i$'s are constants given by
\begin{equation}
C_{1,2,\chi,t}=3/2 \quad \mathrm{and}\quad C_{W,Z,\gamma_L}=5/6,
\end{equation}
and the $n_i$'s are the particle's number of degrees of freedom:
\begin{equation}
n_{W_T}=4, n_{W_L}=n_{Z_T}=2, n_{Z_L}=n_{\gamma_L}=1, n_{1,2}=1, n_{\chi}=3, n_{t}=-12.
\end{equation}
We adopt the method developed by Parwani \cite{parwani1992resummation} to resum the Matsubara zero-modes for the bosonic degrees of freedom. It consists of replacing the bosons' vacuum mass $m_i^2(h,s)$ by the thermal-corrected one $\mathscr{M}_i^2(h,s,T)=m_i^2(h,s)+\Pi_i(T)$, with the self-energy given by 
\bea
\Pi_{s}(T) &=\left(\frac{1}{4} \lambda_{s}+\frac{1}{6} \lambda_{s h}\right) T^{2}, \ \\ 
\Pi_{h}(T) &=\Pi_{\chi}(T)=\left[\frac{1}{16}\left(3 g_{1}^{2}+g_{2}^{2}\right)+\frac{1}{2} \lambda_{h}+\frac{1}{4} y_{t}^{2}+\frac{1}{24} \lambda_{hs}\right] T^{2}, \\ 
\Pi_{W_{L}}(T) &=\frac{11}{6} g_{1}^{2} T^{2}, \\ 
\Pi_{W_{T}}(T) &=\Pi_{Z_{T}}(T)=\Pi_{\gamma_{T}}(T)=0.
\eea
The thermal masses for the longitudinal mode of the photon and $Z$ boson are 
\bea
\mathscr{M}_{Z_L}^2(s, h, T)&=\frac{1}{2}\left[m_Z^2(s, h)+\frac{11}{6} \frac{g_1^2}{\cos ^2 \theta_w} T^2+\Delta(s, h, T)\right]\ \mathrm{and}\\
\mathscr{M}_{\gamma_L}^2(s, h, T)&=\frac{1}{2}\left[m_Z^2(s, h)+\frac{11}{6} \frac{g_1^2}{\cos ^2 \theta_w} T^2-\Delta(s, h, T)\right],
\eea
with
\be
\Delta(s, h, T)=\left[m_Z^4(s, h)+\frac{11}{3} \frac{g_1^2 \cos ^2 2 \theta_w}{\cos ^2 \theta_w}\left(m_Z^2(s, h)+\frac{11}{12} \frac{g_1^2}{\cos ^2 \theta_w} T^2\right) T^2\right]^{1/2}.
\ee\\

At low temperature ($m_i^2/T^2\gg1$), one would expect all the thermal effects to be Boltzmann suppressed, since the species $i$ becomes essentially absent from the plasma. This is manifestly the case for $V_T$, since the thermal integrals decay exponentially in the limit $\mathscr{M}_i^2/T^2\approx m_i^2/T^2\gg1$. However, in the same limit, $V_\mathrm{CW}$ would depend quadratically on $T$ if we used the thermal masses defined above. This would spoil the potential's low-$T$ behaviour. Therefore, we define a regulated thermal mass\footnote{For the photon and $Z$ boson's longitudinal mode, we define $\Pi_i=\mathscr{M}_i^2-m_i^2$, which should reproduce the desired behaviour.} $\tilde{\mathscr{M}}_i^2=m_i^2+R(m_i^2/T^2)\Pi_i$, that should only be used in $V_\mathrm{CW}$. $R(x)$ is a regulator chosen to recover the right behaviour in the low and high-$T$ limit. In order to do so, it should be a smooth function satisfying $R(x=0)= 1$  and $R(x)\sim e^{-\sqrt{|x|}}$ when $|x|\gg1$. We choose here the integrated Boltzmann number density function given by 
\be
R(x)=\frac{1}{2}[x] K_2\left(\sqrt{[x]}\right),
\ee
where $K_2$ is the modified Bessel function of the second kind and $[x]=x \tanh(x)$ is a smoothed absolute value.\\

The last term of Eq.\ (\ref{eq:potential}) contains the following counterterms:
\be
\delta V(h,s) = Ah^2+Bh^4+Cs^2+D,
\ee
which are fixed by requiring the renormalization conditions
\bea
0&=\left. \frac{\partial V_\mathrm{eff}}{\partial h}\right\vert_{h=v,s=0,T=0}\\
m_h^2&=\left. \frac{\partial^2 V_\mathrm{eff}}{\partial h^2}\right\vert_{h=v,s=0,T=0}\\
m_s^2&=\left. \frac{\partial^2 V_\mathrm{eff}}{\partial s^2}\right\vert_{h=v,s=0,T=0}\\
0&=\left. V_\mathrm{eff}\right\vert_{h=v,s=0,T=0}.
\eea
While the use of the resummed one-loop potential is a clear improvement over the leading thermal-mass-corrected approximation, one should keep in mind that higher loop corrections and even nonperturbative physics may be relevant, in particular for very strong transitions~\cite{Brauner:2016fla,Kainulainen:2019kyp,Croon:2020cgk}.

\section{Relativistic fluid equation}
\label{app:fluidEq}

We here calculate the hydrodynamical properties of the plasma close to the wall using the method described in Ref.\ \cite{espinosa2010energy}. The quantities of interest are the temperatures $T_\pm$ and the velocities of the plasma measured in the wall frame $v_\pm$. The subscript $+$ and $-$ indicate that the quantity is measured in front or behind the wall respectively.\\

By integrating the conservation of the energy-momentum tensor equation across the wall, one can show that the quantities $T_\pm$ and $v_\pm$ are related by the equations
\bea \label{eq:fluidEq}
v_+ v_- &= \frac{1-(1-3\alpha_+)r}{3-3(1+\alpha_+)r}, \\
\frac{v_+}{v_-} &= \frac{3+(1-3\alpha_+)r}{1+3(1+\alpha_+)r},
\eea
where $\alpha_+$ and $r$ are defined as
\bea \label{eq:EOS}
\alpha_+ &\equiv \frac{\epsilon_+ - \epsilon_-}{a_+ T_+^4}, \\
r &\equiv \frac{a_+ T_+^4}{a_- T_-^4}, \\
a_\pm &\equiv -\frac{3}{4 T_\pm^3}\left.\frac{\partial V_\mathrm{eff}}{\partial T}\right\vert_\pm, \\
\epsilon_\pm &\equiv \left.\left(-\frac{T_\pm}{4}\frac{\partial V_\mathrm{eff}}{\partial T} + V_\mathrm{eff}\right)\right\vert_\pm.
\eea
These quantities are often approximated by the so-called bag equation of state, which is given in Ref.\ \cite{espinosa2010energy}. This approximation is expected to hold when the masses of the plasma's degrees of freedom are very different from $T$, which is not necessarily true in the broken phase. Therefore, we keep the full relations (\ref{eq:EOS}) in our calculations.\\

Subsonic walls always come with a shock wave in front of the phase transition front. The Eqs.\ \ref{eq:fluidEq} can be used to relate $T_\pm$ and $v_\pm$ at the wall and the shock wave, but we need to understand how the temperature and fluid velocity evolve between these two regions. Assuming a spherical bubble and a thin wall, one can derive from the conservation of the energy-momentum tensor the following differential equations
\bea \label{eq:diffFluidEq}
2\frac{v}{\xi} &= \gamma^2(1-v\xi)\left(\frac{\mu^2}{c_s^2}-1\right)\partial_\xi v, \\
\partial_\xi T &= T\gamma^2\mu\partial_\xi v,
\eea
where $v$ is the fluid velocity in the frame of the bubble's center and $\xi=r/t$ is the independent variable, with $r$ the distance from the bubble center $t$ the time since the bubble nucleation. With that choice of coordinates, the wall is positioned at $\xi=v_w$. $\mu$ is the Lorentz-transformed fluid velocity
\be
\mu(\xi,v) = \frac{\xi-v}{1-\xi v},
\ee
and $c_s$ is the speed of sound in the plasma
\be
c_s^2 = \frac{\partial V_\mathrm{eff}/\partial T}{T\partial^2 V_\mathrm{eff}/\partial T^2} \approx \frac{1}{3}.
\ee
The last approximation is valid for relativistic fluids, which models well the unbroken phase. In the broken phase, the particles get a mass that can be of the same order as the temperature, and it causes the speed of sound to become slightly smaller.\\

One can find three different types of solutions for the fluid's velocity profile: deflagration walls ($v_w<c_s^-$) have a shock wave propagating in front of the wall, detonation walls ($v_w>\xi_J$) have a rarefaction wave behind it and hybrid walls ($c_s^-<v_w<\xi_J$) have both shock and rarefaction waves. $\xi_J$ is the model-dependent Jouguet velocity, which is defined as the smallest velocity a detonation solution can have. Each type of wall have different boundary conditions that determine the characteristics of the solution. Detonation walls are supersonic solutions where the fluid in front of the wall is unperturbed. Therefore, it satisfies the boundary conditions $v_+=v_w$ and $T_+=T_n$. For that type of solution, Eqs.\ (\ref{eq:fluidEq}) can be solved directly for $v_-$ and $T_-$.\\

Subsonic walls always have a deflagration solution with a shock wave at a position $\xi_{sh}$ that solves the equation $v_{sh}^-\xi_{sh} = (c_s^+)^2$, where $v_{sh}^-$ is the fluid's velocity just behind the shock wave measured in the shock wave's frame. It satisfies the boundary conditions $v_-=v_w$ and $T_{sh}^+=T_n$. Because these boundary conditions are given at two different points, the solution of this system can be somewhat more involved than for the detonation case. Indeed, one has to use a shooting method which consists of choosing an arbitrary value for $T_-$, solving Eqs.\ (\ref{eq:fluidEq}) for $T_+$ and $v_+$, integrating Eqs.\ (\ref{eq:diffFluidEq}) with the initial values $T(v_w)=T_+$ and $v(v_w)=\mu(v_w,v_+)$ until the equation $\mu(\xi,v(\xi))\xi=(c_s^+)^2$ gets satisfied. One can then restart this procedure with a different value of $T_-$ until the Eqs.\ (\ref{eq:fluidEq}) are satisfied at the shock wave. Hybrid walls satisfy $v_+ < c_s^- < v_w$ and they have the boundary conditions $v_- = c_s^-$ and $T_{sh}^+ = T_n$, which make them very similar to the deflagration walls.

\section{Gravitational Wave Production}
\label{app:gw}

For the convenience of the reader, we here reproduce the 
formulae from Refs.\ \cite{Caprini:2019egz,Hindmarsh:2017gnf,espinosa2010energy,Guo:2020grp,Hindmarsh:2020hop} that determine the GW spectrum from sound waves and turbulence in a
first order phase transition. The spectrum is~\cite{Guo:2020grp, Hindmarsh:2020hop}
\begin{equation}
\begin{aligned} \label{eq:GW}
\Omega_{\text {sw}}(f) =8.83 \times 10^{-7} K_{\mathrm{sw}}^2\left(\frac{HR}{c_s}\right)\left( 1-\left(1+\frac{2HR}{\sqrt{K_\mathrm{sw}}}\right)^{-1/2}  \right)\left(\frac{100}{g_*}\right)^{1/3} S_{\text {sw }}(f), \end{aligned}
\end{equation}
where $K_\mathrm{sw}=\kappa_\mathrm{sw}\alpha/(1+\alpha)$, with $\kappa_\mathrm{sw}$ the efficiency coefficient of the sound wave. As previously stated, we assume that all the walls have non-runaway solutions and that the contribution from turbulence is negligible; hence we set $\Omega_\mathrm{sw}=\Omega_\phi(f)=0$. The function parametrizing the shape of the GW spectrum is 
\begin{equation}
\begin{aligned} \label{eq:GWspectrum}
S_{\mathrm{sw}}(f) =\left(\frac{f}{f_{\mathrm{sw}}}\right)^{3}\left(\frac{7}{4+3\left(f / f_{\mathrm{sw}}\right)^{2}}\right)^{\frac{7}{2}}, \end{aligned}
\end{equation}
and the peak frequency $f_\mathrm{sw}$ is
\begin{equation}
\begin{aligned}\label{eq:GWfpeak}
f_{\mathrm{sw}}&=2.6 \times 10^{-5}\, \mathrm{Hz}\left(\frac{1}{HR}\right)\left(\frac{T_n}{100\ \mathrm{GeV}}\right)\left(\frac{g_{*}}{100}\right)^{\frac{1}{6}}. \\ 
\end{aligned}
\end{equation}
Numerical fits for the efficiency coefficient $\kappa_{\mathrm{sw}}$ (the fractions of the available vacuum energy that go into kinetic energy) were presented in \citep{espinosa2010energy}. For non-runaway walls, these fits depend on the wall velocity and are given by
\begin{equation}
\kappa_{\mathrm{sw}}= \left\lbrace\begin{matrix}
\frac{c_s^{11/5}\kappa_a\kappa_b}{(c_s^{11/5}-v_w^{11/5})\kappa_b+v_wc_s^{6/5}\kappa_a}, & v_w\lesssim c_s\\ 
\kappa_b+(v_w-c_s)\delta\kappa+\frac{(v_w-c_s)^3}{(\xi_J-c_s)^3}[\kappa_c-\kappa_b-(\xi_J-c_s)\delta\kappa], & c_s<v_w<\xi_J \\
\frac{(\xi_J-1)^3\xi_J^{5/2}v_w^{-5/2}\kappa_c\kappa_d}{[(\xi_J-1)^3-(v_w-1)^3]\xi_J^{5/2}\kappa_c+(v_w-1)^3\kappa_d}, & v_w\gtrsim\xi_J
\end{matrix}\right.
\end{equation}
where $c_s=1/\sqrt{3}$ is the sound velocity and the different parameters are given by 
\begin{equation}
\begin{aligned} 
\xi_J = \frac{\sqrt{2\alpha/3+\alpha^2}+c_s}{1+\alpha} &\hspace{1cm}\delta\kappa=-0.9\log\frac{\sqrt{\alpha}}{1+\sqrt{\alpha}}\\
\kappa_{a} =\frac{6.9 v_w^{6 / 5} \alpha}{1.36-0.037 \sqrt{\alpha}+\alpha} &\hspace{1cm}\kappa_{b} =\frac{\alpha^{2 / 5}}{0.017+(0.997+\alpha)^{2 / 5}} \\
\kappa_c = \frac{\sqrt{\alpha}}{0.135+\sqrt{0.98+\alpha}} &\hspace{1cm}\kappa_d = \frac{\alpha}{0.73+0.083\sqrt{\alpha}+\alpha}
\end{aligned}
\end{equation}

We caution that while these fits, when used as input for a signal-to-noise estimate, are useful to get an overall estimate for the GW signal in a given model, their precise predictions should be interpreted with care. The fit for the sound wave production is reliable for relatively weak transitions $\alpha < 0.1$, which is the range where most of our models fall. For stronger transitions the fit can overestimate the GW-signal by as much as a factor of thousand (strong deflagrations)~\cite{Cutting:2019zws}. In addition to the strength of the transition, fit parameters have also been shown to be sensitive to the shape of the effective potential~\cite{Cutting:2020nla} {\color{black} and the wall velocity \cite{Caprini:2019egz,Hindmarsh:2020hop}.  As explained in Ref.\ \cite{Caprini:2019egz} Eqs.\ (\ref{eq:GW}-\ref{eq:GWfpeak}) are not expected to be accurate for $0.5\lsim v_w\lsim \xi_J$, which includes a large fraction of the deflagration models found in this work.  Thus, pending improvements in the theoretical predictions for GW spectra in this
range of wall speeds, the results should not be regarded as conclusive.}
\\

\end{appendix}
\bibliographystyle{utphys}
\bibliography{cite.bib}
\end{document}